%% file: main.tex
\documentclass[lettersize,journal]{IEEEtran}

\ifCLASSINFOpdf
\else
\fi
\ifCLASSOPTIONcompsoc
  \usepackage[nocompress]{cite}
\else
  \usepackage{cite}
\fi

\usepackage{caption}
\usepackage[none]{hyphenat}
\usepackage{algpseudocode}
\usepackage{amsmath}
\usepackage{amsfonts}
\usepackage{amssymb}
\usepackage[caption=false]{subfig}

\usepackage{graphicx}
\usepackage[table]{xcolor}

\usepackage{array,multirow}

\usepackage{multicol}

\usepackage{mathtools}

\usepackage{amsfonts}
\usepackage{booktabs}
\usepackage{siunitx}

\usepackage{tikz} 
\newcommand*\circled[1]{\tikz[baseline=(char.base)]{%
        \node[shape=rectangle,fill=gray!20,draw,inner sep=2pt,opacity=0.5,text opacity=1] (char) {#1};}}

\newcommand\citem{%
  \stepcounter{enumi}\item[\circled{\arabic{enumi}}]}

\usepackage{enumitem}

\usepackage{url}
\makeatletter
\def\url@leostyle{%
  \@ifundefined{selectfont}{\def\UrlFont{\sf}}{\def\UrlFont{\small\ttfamily}}}
\makeatother
\newcommand{\eat}[1]{}
\usepackage[linesnumbered,ruled]{algorithm2e}
\usepackage{mdframed} 

\renewcommand{\IEEEbibitemsep}{1pt plus 2pt}
\makeatletter
\IEEEtriggercmd{\reset@font\normalfont\small}
\makeatother
\IEEEtriggeratref{1}

\usepackage{xcolor}
\definecolor{light-gray}{gray}{0.9}
\definecolor{light-green}{rgb}{0.133,0.796,0.376}

\newenvironment{packed_enum}{%
  \begin{enumerate}%
  }{\end{enumerate}}

\usepackage{amsthm}
\newtheorem{definition}{Definition}
\newtheorem{theorem}{Theorem}
\newtheorem{lemma}{Lemma}

\newcolumntype{L}[1]{>{\raggedright\let\newline\\\arraybackslash\hspace{0pt}}m{#1}}

\usepackage{listings}

\usepackage{color}
\definecolor{codegreen}{rgb}{0,0.6,0}
\definecolor{codegray}{rgb}{0.5,0.5,0.5}
\definecolor{codepurple}{rgb}{0.58,0,0.82}
\definecolor{backcolour}{rgb}{0.95,0.95,0.92}
\definecolor{shadecolor}{rgb}{0.92,0.92,0.92}
\usepackage{framed} 
 
\lstdefinestyle{mystyle}{
    backgroundcolor=\color{backcolour},   
    commentstyle=\color{codegreen},
    keywordstyle=\color{magenta},
    numberstyle=\tiny\color{codegray},
    stringstyle=\color{codepurple},
    basicstyle=\footnotesize,
    breakatwhitespace=false,         
    breaklines=true,                 
    captionpos=b,                    
    keepspaces=true,                 
    numbers=left,                    
    numbersep=5pt,                  
    showspaces=false,                
    showstringspaces=false,
    showtabs=false,                  
    tabsize=2
}
 
\usepackage{arydshln}

\definecolor{ao}{rgb}{0.0, 0.5, 0.0}

\input{solidity-highlighting.tex}

\begin{document}

\title{T-Watch: Towards Timed Execution of Private Transaction in Blockchains}

\author{
    Chao~Li,~\IEEEmembership{Member,~IEEE,}
    and~Balaji~Palanisamy,~\IEEEmembership{Member,~IEEE,}
}



\maketitle

\begin{abstract}

In blockchains such as Bitcoin and Ethereum, transactions represent the primary mechanism that the external world can use to trigger a change of blockchain state. 
Timed transaction refers to a specific class of service that enables a user to schedule a transaction to change the blockchain state during a chosen future time-frame.
This paper proposes \textsf{T-Watch}, a decentralized and cost-efficient approach for users to schedule timed execution of any type of transaction in Ethereum with privacy guarantees.
\textsf{T-Watch} employs a novel combination of threshold secret sharing and decentralized smart contracts.
To protect the private elements of a scheduled transaction from getting disclosed before the future time-frame, \textsf{T-Watch} maintains shares of the decryption key of the scheduled transaction using a group of executors recruited in a blockchain network before the specified future time-frame and restores the scheduled transaction at a proxy smart contract to trigger the change of blockchain state at the required time-frame.
To reduce the cost of smart contract execution in \textsf{T-Watch}, we carefully design the proposed protocol to run in an optimistic mode by default and then switch to a pessimistic mode once misbehaviors occur. Furthermore, the protocol supports users to form service request pooling to further reduce the gas cost.
We rigorously analyze the security of \textsf{T-Watch} and implement the protocol over the Ethereum official test network. 
The results demonstrate that \textsf{T-Watch} is more scalable compared to the state of the art and could reduce the cost by over 90\% through pooling.

\end{abstract}





\section{Introduction}
\label{sec:introduction}

\IEEEPARstart{B}{lockchains}
are distributed ledgers of transactions performed on a global state by nodes of blockchain networks.
Transactions are at the heart of blockchains because they are the only things that the external world can use to trigger a change of blockchain state.
In Bitcoin-like blockchains~\cite{nakamoto2008bitcoin}, transactions allow users to transfer funds among each other.
In Ethereum-like blockchains~\cite{buterin2014next} that support smart contracts~\cite{wood2014ethereum}, transactions also enable users to deploy and interact with smart contracts.
Recent advancements in blockchain technology have led to a proliferation of transactions happening in blockchain networks. 
In 2021, the median number of daily Bitcoin and Ethereum transactions has grown to 270,736
 and 1,245,624, respectively~\cite{bitinfocharts}.

\underline{T}imed \underline{T}ransaction (\textit{TT}) refers to a class of service that enables a user to schedule a transaction to get executed to change the blockchain state during a chosen future time-frame.
Many scenarios require timed transactions in blockchains. 
For example, just like the clients of credit card and facility companies in real-world who are usually allowed to select a predetermined time-frame during which payments are charged from their bank accounts, the clients of vendors or service providers in blockchain networks also need a way to schedule payment transactions to transfer cryptocurrencies from their blockchain accounts.
For many businesses and computations running over smart contracts, the ability of \textit{TT} to control the execution time of sensitive transactions and knowing precisely when such transactions are executed can be crucial.
For instance, imagine that Alice is working on a confidential project and she needs to outsource a computation task to service providers by sending out a transaction that creates a smart contract implementing verifiable cloud computing~\cite{dong2017betrayal}, she may only want the contract to be recorded into the ledger exactly during the acceptance phase of the project because an early created contract could potentially leak her rate of progress to her competitors.
Imagine another situation in which Bob attends a smart-contract-based sealed-bid auction~\cite{SealedBidAuction} and would like his bid to get revealed via a transaction calling the \textit{reveal()} function at the auctioneer smart contract exactly during the bid opening time-frame. Here, an early revelation could potentially leak his bid to his competitors who may have kept tracking Bob's transasctions while a late revelation will remove Bob's bid from the competition.

While there are numerous services (e.g., BlueOrion~\cite{blueorion} and Oraclize~\cite{oraclize}) that provide pre-scheduled execution of \textit{TT} in blockchains, most of current implementations of \textit{TT} are highly centralized,
where users are forced to fully trust the centralized service providers, resulting in various security issues induced by a single point of trust.
Moreover, even in cases where the service providers are perceived as trustworthy, the services remain vulnerable to unpredictable security breaches or internal attacks over which they have no control~\cite{chen2018certchain}.
Recently, the emergence of blockchain technologies offers great potential to decentralize implementations of \textit{TT}, 
and there have been several proposals of decentralized implementations~\cite{Clock,ning2018keeping,li2018decentralized-2,Kimono,SilentDelivery}.
Nevertheless, these implementations for now only support a
single type of transaction in Ethereum, namely function invocation transaction.
More importantly, most of the smart-contract-based protocols developed in existing decentralized approaches can hardly become practical because they typically involve $O(n)$ cost of running smart contracts among $n$ protocol participants while they need a larger $n$ to maintain higher security.

In this paper, we present \textsf{T-Watch}, a decentralized implementation of \textit{TT} for users to schedule \textit{any} type of transaction in Ethereum with protection of private data within scheduled transactions with significantly reduced cost of running smart contracts during the process.  
\textsf{T-Watch} employs a novel combination of threshold secret sharing and decentralized smart contracts.
To protect the private elements of scheduled transactions from getting disclosed before the future time-frame, 
\textsf{T-Watch} maintains shares of the decryption key of the scheduled transaction signed by a user using a group of executors recruited in a blockchain network before the specified future time-frame and restores the scheduled transaction at a proxy smart contract to trigger the change of blockchain state exactly during the required time-frame.
To reduce the cost of running smart contracts in each timed transaction, 
we carefully design a protocol to run in an optimistic execution path called \textit{T-Opt} by default, where the non-scalable regulations are cut off to reduce the cost of running smart contracts.
Then, upon detecting any misbehavior, executors reserve the ability to switch the protocol to a pessimistic execution path called \textit{T-Pes} by rebinding the removed regulations with smart contracts to redress and penalize any dishonest behavior, just as if these regulations were never decoupled. 
Finally, the protocol encourages users to form service request pools through an execution path called \textit{T-Pool} so that the shares of a certain decryption key could be shared among multiple users who may have the same required time-frame.
The proposed protocol incentivizes executors to stay honest and thus implements \textsf{T-Watch} with minimum interactions on the blockchain.


In summary, we make the following key contributions:
\begin{itemize}[leftmargin=*]
\item To the best of our knowledge, \textsf{T-Watch} is the \textit{first} practical decentralized approach designed for \textit{TT} that is secure, scalable, privacy-preserving and cost-efficient. 
\item After the service has been set up, \textsf{T-Watch} completely isolates the service execution from the state of users, without requiring any assistance from the user side.
\item We emphasize that \textsf{T-Watch} is a general approach that supports \textit{all} types of transaction in Ethereum.
\item \textsf{T-Watch} can protect the privacy of \textit{all} elements of all types of transaction before a prescribed time-frame, including the amount of transferred funds, the content of created smart contracts, the arguments of invoked functions and even the addresses of payees or invoked smart contracts.
\item We carefully design a protocol that could implement \textsf{T-Watch} through three different execution paths, an optimistic path \textit{T-Opt} that spends a limited amount of gas in executing the prescribed transaction if no participant misbehaves,
a pessimistic path \textit{T-Pes} that is relatively gas-consuming but could resist misbehaviors, 
and a service request pooling path \textit{T-Pool} that reduces the cost of most service requests in pools to a very small and constant value.
\item We rigorously analyze the security of \textsf{T-Watch} against two different threat models, $\mathcal{A}$-threshold and $\mathcal{A}$-budget. 
We implement the protocol over the Ethereum official test network. 
The results demonstrate that \textsf{T-Watch} is more scalable compared to the state of the art and could reduce the gas cost by over 90\% through pooling.  
\end{itemize}

\textcolor{black}{While our focus is primarily on Ethereum, the versatile design of \textsf{T-Watch} enables its adaptation to numerous EVM-compatible blockchains such as Binance Smart Chain, Tron, and Polygon. 
Additionally, by abstracting the core functionalities of \textsf{T-Watch} and adapting the transaction handling procedures, it can be made compatible with a variety of other blockchains that might not directly support EVM.}

\section{Related work}
\label{s2}


\subsection{Timed transaction in blockchains}
\label{s2.1}

Existing techniques and tools for timed transaction in blockchains can be roughly divided into two categories:

\noindent \textbf{Centralized approaches}:
Recently, many emerging blockchain services companies such as BlueOrion~\cite{blueorion} and Oraclize~\cite{oraclize}  offer services of scheduling \textit{TT}.
These services are heavily centralized and require users to trust the companies.
Alternatively, users may themselves run client-side tools such as \textit{parity}~\cite{parity} to schedule \textit{TT}.
Nevertheless, such tools typically fail to isolate users from the service execution because users have to make their machines keep connecting with the blockchain network.

\noindent \textbf{Decentralized approaches}:
In Bitcoin, there exists a native mechanism named \textit{Timelocks}~\cite{antonopoulos2017mastering} that allows users to set a timelock for a payment transaction so that the Bitcoin carried by that transaction becomes available only after the specified time point.
Unfortunately, \textit{Timelocks} is proprietary to Bitcoin and is not supported by most other blockchains such as Ethereum, so it cannot support any other type of transactions or protect sensitive elements of scheduled transactions.
Recently, a project named \textit{Ethereum Alarm Clock}~\cite{Clock} proposes to recruit Ethereum accounts to trigger a re-deployed contract to call a target contract during a prescribed time-frame.
However, this scheme neither protects sensitive elements nor guarantees the transaction execution.
A more recent work~\cite{li2018decentralized-2} further leverages threshold secret sharing~\cite{shamir1979share} to offer a certain level of protection of sensitive elements, but it only supports function invocation transactions and has a cost of $O(n)$.

\subsection{Timed release of private data}
\label{s2.2}

We design \textsf{T-Watch} to protect sensitive elements within all types of timed transactions before the prescribed time-frame, which is relevant to a classical research topic, namely the timed release of private data.
The study of timed release of private data began with May~\cite{may1992timed}.
Since then, there have been extensive studies on this problem, which can be roughly divided into four categories.

\noindent \textbf{Mathematical puzzles}: 
One representative approach~\cite{rivest1996time,mahmoody2011time} protects private data with a time-lock puzzle, forcing recipients to solve a cryptographic puzzle to obtain the data. 
Nevertheless, the time for solving such puzzles is non-deterministic and hence, the opening time of the data can not be precisely controlled. 
Also, cryptographic approaches for timed data release come with a very significant computational cost and as such, these techniques are not scalable.

\noindent \textbf{Time server}: 
Another well-studied approach~\cite{emura2011timed,kasamatsu2012time} relies on a (semi-)trusted time server to release time trapdoors to recipients at specified future time points.  
These techniques involve a single point of trust and create a safety bottleneck.

\noindent \textbf{Proof-of-Work puzzles}: 
One direction of the decentralized approaches~\cite{liu2018build,liu2015time} encloses private data with blockchain puzzles used in Proof-of-Work~\cite{nakamoto2008bitcoin} and therefore minimizes the computational burden of the data recipients as the blockchain puzzles are periodically solved by blockchain miners.
However in such an approach, the involved heavy cryptographic primitives result in very high performance overhead~\cite{liu2015time,ning2018keeping}. 

\noindent \textbf{Smart contracts}: 
Another direction of recent decentralized techniques for timed data release~\cite{ning2018keeping,Kimono,SilentDelivery} leverages smart contracts to establish a decentralized autonomous agent, through which a data sender could recruit a group of peers from the blockchain network as her trustees to cooperatively maintain and deliver her private data to recipients. 
While the above-mentioned work considered a rational adversarial setting, the issue of protecting timed-release services in mixed adversarial environments consisting of both malicious and rational peers is addressed in \cite{icbc2022, dbsec2022}. 

We emphasize the problem of timed transactions is more general and challenging than timed data release in the context of smart contracts. In timed data release, data is released to recipients who can actively leverage their accounts to interact with other participants of the protocol and exchange messages via private off-chain communication channels.
In contrast, recipients of data carried by timed transactions could also be smart contract accounts that are both passive and transparent, so solutions for smart-contract-based timed data release are not capable of handling timed transactions.
Besides, timed transactions enable new applications that are not supported by timed data release, including timed creation of smart contracts and timed payments of cryptocurrencies.

To sum up, our work in this paper tackles the key limitations of the state-of-the-art decentralized approaches~\cite{Clock,ning2018keeping,Kimono,li2018decentralized-2,SilentDelivery}. To the best of our knowledge, \textit{T-Watch} is the \textit{first} practical decentralized solution for timed transaction with strong guarantees of security, scalability and cost-efficiency.

%

\section{Preliminaries}

\subsection{Account types in Ethereum} 
There are two types of accounts in Ethereum:
\begin{itemize}[leftmargin=*]
  \item External Owned Account (EOA): To interact with the Ethereum ecosystem, one needs to own an EOA by locally creating a pair of asymmetric keys, leveraging the public key to generate a unique address \textit{addr(EOA)} for this EOA and preserving the private key to sign future transactions.
  \item Contract Account (CA): Smart contracts in Ethereum are typically created through contract creation transactions. Each new smart contract is automatically assigned a unique address \textit{addr(CA)} in a deterministic way.
\end{itemize}



\subsection{Transactions and update patterns}
\label{s3.2}
A transaction in Ethereum is a serialized binary message sent from an EOA that contains the following key elements:
\begin{equation}
\label{e1}
\textit{tx}\ \equiv\ \langle 
\textit{to},\textit{value},\textit{data},\textit{sig} \rangle
\end{equation}
\begin{itemize}[leftmargin=*]
  \item \textit{tx.to}: the account address of the recipient (EOA or CA).
  \item \textit{tx.value}: the amount of \textit{ether}
  to send to the recipient.
  \item \textit{tx.data}: the binary data carried by \textit{tx}.
  \item \textit{tx.sig}: the ECDSA digital signature of the EOA.
\end{itemize}

Depending on the type of \textit{tx.to}, transactions can be divided into three categories.

\noindent \textbf{Fund transfer transaction}:
In Ethereum, to transfer an amount of cryptocurrency \textit{ether} to another EOA account, one needs to create a fund transfer transaction by setting the recipient EOA as \textit{tt.to} as well as a non-empty \textit{tt.value}. 


\noindent \textbf{Function invocation transaction}: 
To invoke a function within a deployed smart contract, one needs to create a function invocation transaction by setting the CA (contract account) as \textit{tt.to} as well as a non-empty \textit{tt.payload}.

\noindent \textbf{Contract creation transaction}: 
To create a new smart contract, one needs to create a contract creation transaction by setting 0x0 as \textit{tx.to} , an empty \textit{tx.value} and a non-empty \textit{tx.data}.
Here, the created new smart contract shall be programmed using a high-level contract-oriented language such as \textit{Solidity}~\cite{Solidity2017} and then compiled into a low-level bytecode language called Ethereum Virtual Machine (EVM) code.
Finally, the contract creation transaction carries the bytecode as its \textit{tx.data}.

A user, after filling the elements, will then sign the transaction with the private key of a controlled EOA to obtain \textit{tx.sig}.
After that, following the Proof-of-Work (PoW) consensus protocol~\cite{nakamoto2008bitcoin,wood2014ethereum}, the created transaction is broadcasted to the Ethereum network, executed by miners and finally packed into the blockchain ledger.
In Ethereum, transactions and smart contracts are executed transparently by tens of thousands of miner nodes in a decentralized manner and the results are deterministic.

\begin{figure}
\centering
{
   
    \includegraphics[width=9cm,height=1.35cm]{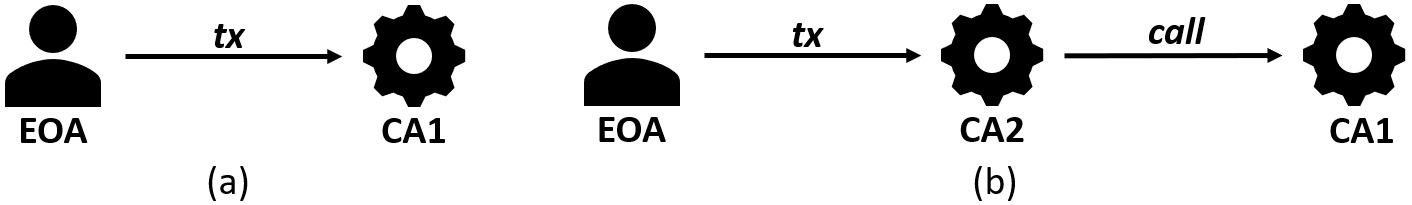}
}
\vspace{-5mm}
\caption {EOA-only pattern (a) and EOA-CA pattern (b) to update the state of blockchain}
\vspace{-5mm}
\label{preliminary} 
\end{figure}

\noindent \textbf{Patterns to update the state of blockchain}:
The state of Ethereum blockchain could be updated via two patterns:
\begin{itemize}[leftmargin=*]
  \item \textbf{EOA-only pattern}: One may leverage EOAs to directly transfer funds, invoke functions or create smart contracts by establishing transactions accordingly.
  For instance, in Fig.~\ref{preliminary}.(a), the EOA creates a function invocation transaction to directly invoke a target function at CA1.
  \item \textbf{EOA-CA pattern}: Alternatively, one may leverage EOAs to create function invocation transactions only to invoke special functions at CAs, where EVM-level opcodes for system operations, such as \texttt{CALL} and \texttt{CREATE}, have been enabled. 
  Then, the opcode \texttt{CALL} allows a contract to transfer funds to another contract or invoke a function at another contract and the opcode \texttt{CREATE} allows an existing contract to deploy a new contract.
  In Fig.~\ref{preliminary}.(b), the EOA creates a transaction to invoke a function at CA2, where the opcode \texttt{CALL} further triggers the invocation of the target function at CA1.
  Here, it is worth noting that CAs alone are unable to update the blockchain state because the execution of any function must be triggered by transactions created at EOAs from the root.
\end{itemize}

\subsection{Gas system}
\label{s3.3}

It is worth noting that the execution of transactions in Ethereum costs gas, which in turn generates transaction fees.
Concretely, each transaction in Ethereum first charges a basic fee of 21,000 Gas. 
On this basis, each instruction in the transaction execution process is charged the corresponding gas fee according to its type~\cite{wood2014ethereum}.
Depending on the complexity of the transaction execution, the total amount of gas spent in a single transaction can be in the millions, but cannot exceed the maximum amount of gas allowed in a single block. 
In Ethereum, in order to send an executable transaction, one needs to ensure that the balance of the sender EOA is sufficient. 
This is because 
the \textit{ether} corresponding to the total amount of gas spent by the transaction will be charged.
The gas system is important as it helps incentivize miners to remain honest, suppress denial-of-service attacks, and encourage efficient smart contract programming. 
However, the gas system puts forward higher requirements for the scalability of the protocol design. In a protocol with many participants, even if the cost of a single transaction is low, the total fee of all the transactions may be very high.
\subsection{Off-chain channels}
\label{s3.4}
Like most blockchain systems, nodes in Ethereum form a peer-to-peer (P2P) network. 
The Ethereum community proposed the Whisper protocol~\cite{Whisper2017} to support inter-node communication in the P2P network. 
Specifically, messages sent based on the Whisper protocol are broadcast over the entire P2P network, and all messages must be encrypted symmetrically or asymmetrically and can be decrypted by the node with the corresponding key. 
For example, a node can generate a pair of asymmetric keys locally and store the public key on the blockchain, making the public key visible to the entire network, so as to obtain encrypted messages transmitted by other nodes using the public key and the Whisper protocol, and decrypt these messages to learn the contents.
Unlike on-chain transactions that charge fees, off-chain communication costs no money.

\subsection{Cryptographic tools}
\label{s3.5}
\textsf{T-Watch} employs several key cryptographic tools:
\begin{itemize}[leftmargin=*]
  \item \textbf{Threshold secret sharing}:
  A $(t,n)$-threshold secret sharing scheme is a method to split a secret $s$ into $n$ shares $\boldsymbol{s}=\{s_1,...,s_n\}$ such that any $t$ shares could recover the secret but $t-1$ or fewer shares fail to do that. 
  This paper adopts the scheme proposed by Shamir based on Lagrange interpolation theorem~\cite{shamir1979share}, and denote the secret split and secret restoration algorithms as 
  $\boldsymbol{s} \gets \textit{SS}(s,\{t,n\})$ and $s \gets \textit{SR}(\boldsymbol{s},\{t,n\})$, respectively.
  \item \textbf{Keccak-256 hash function}: 
  Keccak-256~\cite{bertoni2013keccak} has been widely used in blockchain systems including Ethereum. 
  The Ethereum Solidity language provides a function $keccak256(\cdots)\ returns\ (bytes32)$, which can directly calculate Keccak-256 hash values in smart contracts. 
  All hash functions used in this paper are based on Keccak-256, and the hash operation is expressed as $h \gets \textit{H}(\cdot)$.
  \item \textbf{ECDSA}: 
  Many mainstream blockchain systems such as Ethereum and Bitcoin adopt Elliptic Curve Digital Signature Algorithm (ECDSA)~\cite{johnson2001elliptic} for their public-key cryptography. 
  Concretely, in Ethereum, one needs to first randomly generate a private key $sk$ that satisfies the order of the secp256k1 curve and then generate the corresponding public key $pk$ from the private key.
  In the rest of the paper, 
  we simply denote the corresponding encrytion and decryption algorithms as 
  $c \gets \textit{E}(pk,p)$ and $p \gets \textit{D}(sk,c)$, respectively.
  Besides, an ECDSA signature consist of two integers $\{ r,s \}$, upon which Ethereum introduces an additional recovery identifier $v$, forming a triplet signature $\{ v,r,s \}$. 
  The Ethereum community provides a JavaScript API to sign arbitrary messages $msg$, and the signature operation is expressed as $vrs \gets \textit{S}(\textit{H}(msg))$.
  In addition, the Ethereum Solidity language provides a function $ecrecover(\cdots)\ returns (address)$, which can directly verify the signature in smart contracts, represented as 
  $addr(signer) \gets \textit{V}(\textit{H}(msg),vrs)$, and the output $addr(signer)$ is the signer’s EOA address.
  \item \textbf{ECVRF}:
  The Elliptic Curve Verifiable Random Function (ECVRF)~\cite{vrf} is a VRF that uses Elliptic Curves.
  Concretely, it allows a prover holding a keypair $\langle  pk,sk \rangle$ from EC (e.g., secp256k1) to leverage the private key $sk$ and an input message $msg$ to create a pseudo-random number $r \gets \textit{VR}(sk,msg)$, as well as a corresponding proof $\pi \gets \textit{VP}(sk,msg)$.
  Then, the proof $\pi$ allows a verifier holding the public key $pk$ to verify that $r$ is valid under $msg$ and $pk$ through $r \gets \textit{VV}(pk,msg,\pi)$. 
\end{itemize}



\section{T-Watch: Overview}
In this section, we provide an overview of \textsf{T-Watch}, with the intuition behind our core techniques.
We first present the architectures for sending timed transactions.
We then briefly depict the potential execution paths in \textsf{T-Watch} and introduce the core idea that reduces the on-chain cost. 
We defer detailed protocols fulfilling these execution paths until Section~\ref{s5}.
We finally describe the threat models in \textsf{T-Watch}.



\subsection{Architectures for sending timed trasanctions}
\label{s4.1}

Recall that a transaction consists of the components shown in Eq.~\ref{e1} and any state change 
must be triggered by EOAs from the root,
we can abstract the following four key components from an architecture for sending timed trasanctions ($tt$):
\begin{equation}
\label{e2}
\textit{tt}\ \equiv\ \langle  \textit{timer},\textit{EOA},\textit{payload},\textit{fund} \rangle
\end{equation}
\begin{itemize}[leftmargin=*]
  \item $tt.timer$: It expresses the future time-frame (e.g., start and end block numbers) scheduled for releasing $tt$.
  \item $tt.EOA$: It indicates the EOA who triggers the release of $tt$ during the prescribed time-frame.
  \item $tt.payload$: 
  It refers to the payload data carried by $tt$, which needs to be maintained until $tt.timer$ to be sent with $tt$. Here, $tt.payload$ includes the following three elements of a scheduled transaction $tx$:
  \begin{equation}
  \textit{tt.payload}\ =\ \langle  \textit{tx.to},\textit{tx.value},\textit{tx.data} \rangle
  \end{equation}
  \item $tt.fund$: 
  It denotes an amount of $ether$ prepared to satisfy the amount indicated by $tx.value$.
  Specifically, $|tt.fund| = |tx.value|$, and $tt.fund$ needs to be maintained and made available during the time-frame indicated by $tt.timer$ to be transferred with $tt$. 
\end{itemize}


\begin{figure}
\centering
{
   
    \includegraphics[width=1\columnwidth]{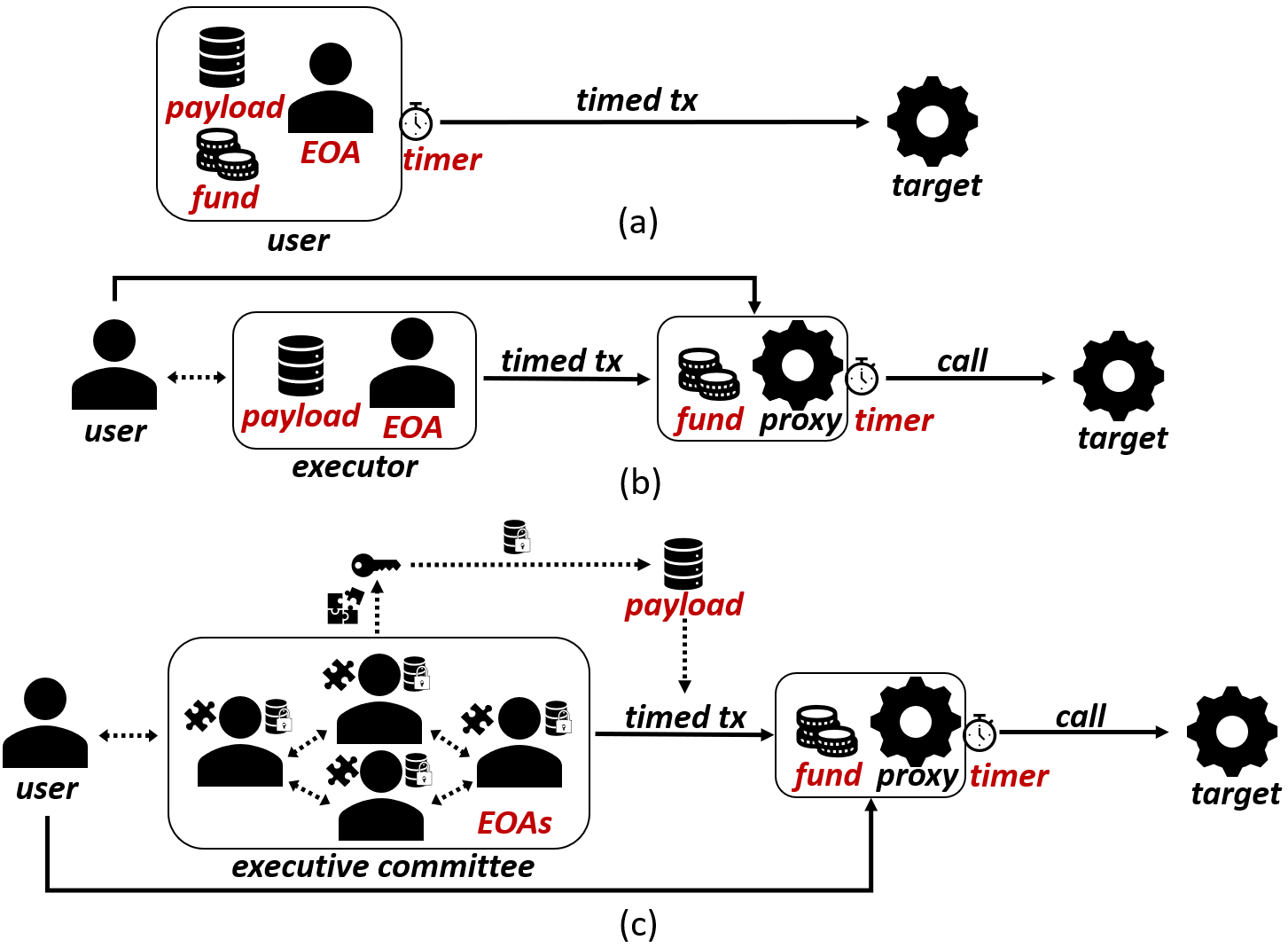}
}
\vspace{-4mm}
\caption {The user-driven architecture (a), the executor-driven architecture (b), and the committee-driven architecture (c) for sending timed transactions.
The solid/dotted lines indicate on-chain/off-chain operations, respectively.}
\vspace{-4mm}
\label{architecture}
\end{figure}

After abstracting the four key components, based on different design options for implementing them, we next present three architectures for sending timed transactions by using examples shown in Fig.~\ref{architecture}, where the goal is to invoke a target smart contract during a prescribed time-frame.

\noindent \textbf{User-driven architecture}:
A naive design option is to adopt the EOA-only pattern presented in Section~\ref{s3.2} to invoke the target contract, which requires users of timed transaction service to handle all the four key components by themselves.
As shown in Fig.~\ref{architecture}.(a), in this architecture, a user needs to keep connecting with the Ethereum network, maintain $tt.payload$ locally with a timer $tt.timer$, and leverage a controlled EOA with sufficient balance to afford both $tt.EOA$ and $tt.fund$.
Then, during the time-frame indicated by $tt.timer$, client tools such as \textit{parity}~\cite{parity} could help send out the scheduled transaction carrying both $tt.payload$ and $tt.fund$ from the EOA.
However, the user-driven architecture cannot isolate service execution from the user side after the service has been set up, making it difficult to be adopted as a general approach.

\noindent \textbf{Executor-driven architecture}:
To completely isolate the service execution from the user side, 
a natural design option is to make users recruit experienced people from the Ethereum community as executors, who may serve timed transactions with more professional equipment.
Here, a naive choice is to outsource all the four key components to recruited executors so that they could simply follow the user-driven architecture on behalf of users.
However, despite that the integrity of \textit{tt.payload} could be verified using users' signatures, outsourcing \textit{tt.fund} and \textit{tt.timer} directly to EOAs controlled by executors is insecure because a dishonest executor may easily embezzle \textit{ether} received from a user and may also violate the indication of \textit{tt.timer}.
Therefore, as shown in Fig.~\ref{architecture}.(b), we adopt the EOA-CA pattern presented in Section~\ref{s3.2} to replace the EOA-only pattern so that a \textit{proxy} contract deployed by a user can act on behalf of the user to preserve \textit{tt.fund} and enforce \textit{tt.timer} with codes in a trustworthy and decentralized manner.
However, the rest two components \textit{tt.payload} and \textit{tt.EOA} remain to be managed in a centralized way and thus become the bottleneck of the architecture because a malicious recruited executor may either disclose or abuse sensitive \textit{tt.payload} or be derelict of its duty of being \textit{tt.EOA}.

\noindent \textbf{Committee-driven T-Watch architecture}:
Therefore, as shown in Fig.~\ref{architecture}.(c), we further extend the executor-driven architecture to the committee-driven \textsf{T-Watch} architecture that can completely decentralize all the four components and also prevent \textit{tt.payload} from getting disclosed before the prescribed time-frame.
The key idea behind the \textsf{T-Watch} architecture is to recruit a group of executors to form an executive committee and jointly maintain \textit{tt.payload} via the threshold secret sharing~\cite{shamir1979share}.
More specifically, a user generates a keypair $\langle pk_u,sk_u \rangle$, encrypts \textit{tt.payload} with $pk_u$ and splits $sk_u$ to shares $\{s_1,...,s_n\}$. 
Then, each share $s_i$ is maintained by one or multiple executors before the specified time-frame.
During the time-frame, the executors jointly restore $sk_u$ and call the \textit{proxy} contract with the decrypted \textit{tt.payload}.
With \textsf{T-Watch}, as long as a certain amount of recruited executors stay honest, \textit{tt.payload} shall be concealed before the time-frame while revealed during the time-frame.
Besides, as long as a single recruited executor is honest, the task of \textit{tt.EOA}, namely calling the \textit{proxy} contract shall be completed during the time-frame.
Therefore, we say that this architecture is able to completely decentralize all the components and also protect the content of all the components within \textit{tt.payload}, including \textit{tx.to}, \textit{tx.value} and \textit{tx.data}.
\textcolor{black}{Besides, we believe that shifting the assumption from the honesty of a single executor to the honesty of some executors makes it more suitable for the blockchain context.}

\noindent \textbf{On/off-chain operations}:
The operations performed by users and executors in Fig.~\ref{architecture} consists of two modes widely recognized by relevant studies~\cite{maram2019churp,kosba2016hawk}.
\begin{itemize}[leftmargin=*]
  \item \textit{On-chain operations}: 
  As presented in Section~\ref{s3.2}, participants can make data publicly recorded by the blockchain using function invocation transactions that carry the data in \textit{tx.data}.
  We assume consistency, availability, and immutability of blockchain systems. 
  Specifically, when a participant submits a message to the blockchain through a transaction, within a limited period of time, other participants can obtain the message from the ledger, and the content is consistent and immutable.
  \item \textit{Off-chain operations}: 
  As presented in Section~\ref{s3.4}, participants can establish P2P off-chain channels using the Whisper protocol to deliver messages to specific recipients.
  We assume availability and reliability of off-chain channels. 
  We also assume that off-chain channels conform to the synchronous model, that is, there is a definite upper bound on the delay of transmitting messages.
\end{itemize}

\subsection{Execution paths of \textsf{T-Watch}}
\label{s4.2}
We consider three execution paths of \textsf{T-Watch} with different characteristics, namely
the optimistic path \textit{T-Opt},
the pessimistic path \textit{T-Pes},
and the service request pooling path \textit{T-Pool}:
\begin{itemize}[leftmargin=*]
  \item \textbf{\textit{T-Opt} (optimistic)}: 
  This optimistic path is the default option and it assumes that there would be no misbehaviors. 
  In short, we expect \textit{T-Opt} to incur $O(n)$ gas cost for a user to schedule a timed transaction $tt$ by establishing a new committee and only $O(1)$ gas cost for the committee to execute the prescribed $tt$ if no participant misbehaves.
  \item \textbf{\textit{T-Pes} (pessimistic)}: 
  As the back-up option, the \textit{T-Pes} path is pessimistic, which is gas-consuming but could resist misbehaviors. 
  The path is designed to be activated to replace \textit{T-Opt} only at the moment when misbehaviors occur.
  It could resist misbehaviors at the expense of $O(n)$ gas cost for the committee to execute the prescribed timed transaction $tt$.
  \item \textbf{\textit{T-Pool} (service request pooling)}: 
  For a user (say Bob) who may not want to afford the $O(n)$ gas cost of setting up a committee, \textit{T-Pool} allows Bob to join a service request pool as a follower.
  The leader of the service request pool is another user (say Alice), whose committee would later execute Bob's $tt$ at the same time-frame prescribed by Alice along with her $tt$, thus reducing Bob's cost from $O(n)$ to $O(1)$.
  In each service request pool, all the followers directly inherit the same level of resistance towards misbehaviors from the leader.
  Therefore, \textit{T-Pool} also relies on \textit{T-Pes} to handle potential misbehaviors.
\end{itemize}

\begin{figure}
\centering
{
   
    \includegraphics[width=1\columnwidth]{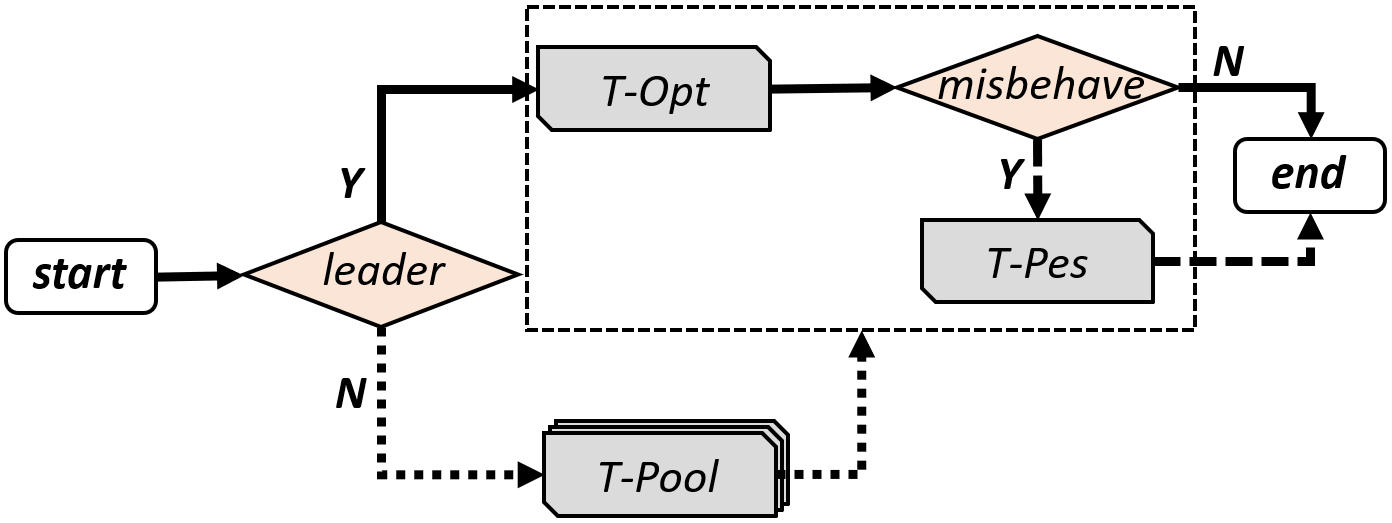}
}
\vspace{-4mm}
\caption {
A flowchart of the three execution paths of \textsf{T-Watch}.
}
\vspace{-4mm}
\label{path}
\end{figure}

Next, we present the strategy for selecting and switching execution paths using a flowchart shown in Fig.~\ref{path}.
At the outset, 
a user needs to make a choice between \textit{T-Opt} and \textit{T-Pool}.
Here, a user employing \textit{T-Opt} would become a leader who initializes a new service request pool of $tt$ while a user who has chosen \textit{T-Pool} would become a follower to join an existing pool.
It is worth noting that a follower should only join a service request pool if the future time-frame prescribed by the leader meets the demand of the follower. 
After that, in \textit{T-Opt}, if no participant misbehaves before the prescribed time-frame, the timed transaction $tt$ will be released during the time-frame and the entire process will complete with \textit{T-Opt}.
Otherwise, \textit{T-Opt} will be switched to \textit{T-Pes} so that misbehaviors can be appropriately addressed and the process will complete with \textit{T-Pes}.
Besides, in \textit{T-Pool}, depending on whether misbehaviors occur or not, \textit{T-Pool} will be switched to \textit{T-Pes} or \textit{T-Opt}, respectively, and the process will complete with \textit{T-Pes} or \textit{T-Opt}, correspondingly.

\subsection{Threat Model and Assumptions}
\label{s4.3}

Recent works~\cite{dong2017betrayal,li2020nf} on blockchain-based protocol design suggest four different types of participants of a protocol, namely honest (never violate a protocol), semi-honest, malicious and rational participants:
\begin{itemize}[leftmargin=*]
  \item \textbf{\textit{Semi-honest participants}}: 
  They behave according to the protocol but try to obtain more information from the available intermediate results.
  \item \textbf{\textit{Malicious participants}}: 
  They can take any malicious actions without regard to their own interests.
  \item \textbf{\textit{Rational participants}}: 
  The behavior of this type of participant is driven by self-interest.
  When the expected payoff from following a protocol is higher, the protocol will be followed, otherwise, the protocol will be violated.
\end{itemize}

Recent studies~\cite{groce2012fair,nguyen2013analyzing} believe that in many practical situations, the assumption of semi-honest participants is too weak, while that of malicious participants is too strong. Therefore, the assumption of rational participants driven by personal interests is considered to be more realistic in many attack scenarios in blockchain~\cite{dong2017betrayal,li2020nf,kumaresan2014use,das2018yoda}. 
In this paper, we assume that there exists an adversary $\mathcal{A}$ seeking the premature disclosure of private \textit{tt.payload} before the future time-frame. 
For instance, Bob's competitors in a sealed-bid auction may want to obtain the \textit{tt.payload} from Bob's $tt$ to learn Bob's bid before the bid opening time-frame.
It is easy to see that such premature disclosure could happen only when $\mathcal{A}$ acquires at least $t$ shares out of a total of $n$ shares of the decryption key before the time-frame.

We present two attack methods that $\mathcal{A}$ may mount to disclose the protected \textit{tt.payload} before the prescribed time-frame in the context of \textsf{T-Watch}:
\begin{itemize}[leftmargin=*]
  \item \textbf{\textit{Sybil attack}}: 
  With Sybil attacks~\cite{douceur2002Sybil}, $\mathcal{A}$ can create a large number of EOAs and use them to occupy as many as positions of recruited executors of a targeted user.
  Consequently, immediately after the shares are assigned to the recruited executors, it is possible for $\mathcal{A}$ to restore the decryption key and obtain \textit{tt.payload} in plaintext.
  \item \textbf{\textit{Bribery attack}}: 
  Instead of occupying the positions of recruited executors in person, $\mathcal{A}$ may choose to prepare a fund and use the fund as a reward to bribe executors recruited by a targeted user that are not controlled by $\mathcal{A}$.
  In addition, $\mathcal{A}$ could even create a bribery smart contract to establish fair collusion with a disloyal executor recruited by the targeted user~\cite{dong2017betrayal}.
\end{itemize}

We understand that it is difficult to completely prevent these attacks in public blockchains, especially when $\mathcal{A}$ has unlimited power.
This paper considers two different threat models:

\noindent \textbf{$\mathcal{A}$-threshold}:
Inspired by the threat models adopted in~\cite{wust2020ace,maram2019churp}, $\mathcal{A}$-threshold assumes that $\mathcal{A}$ is malicious, but could corrupt no more than $tl-1$ executors through either Sybil or Bribery attack in a committee consisting of $nl$ executors, where $t$ shares are generated through $(t,n)$-threshold secret sharing and each share is jointly maintained by $l$ executors. 
Besides, $\mathcal{A}$-threshold assumes that users and the remaining executors are honest.
Under this model, we define a $\mathcal{A}$-threshold-resistant protocol as follows:
\label{definition1}
\begin{definition}
    A $\mathcal{A}$-threshold-resistant protocol 
    satisfies the secrecy property: \\
    \textit{\textbf{Secrecy}}: 
    If $\mathcal{A}$ corrupts no more than $tl-1$ executors in the committee, $\mathcal{A}$ learns no extra information about $sk_u$. 
\end{definition}

\noindent \textbf{$\mathcal{A}$-budget}:
Recent efforts~\cite{dong2017betrayal,matsumoto2017ikp} frequently leverage cryptocurrency (e.g., \textit{bitcoin}, \textit{ether}) as security deposits to incentivize participants to obey protocols.
Inspired by them, 
$\mathcal{A}$-threshold assumes that $\mathcal{A}$ is malicious, but has a bounded attack budget denoted as $b$, namely a bounded total amount of \textit{ether} that can be invested in either Sybil or Bribery attack. 
Besides, $\mathcal{A}$-threshold assumes that all the participants, including executors and users, are rational. 
Under this model, we denote the prescribed amount of security deposit per service per executor as $\Delta d$ and define a $\mathcal{A}$-budget-resistant protocol as follows:
\label{definition2}
\begin{definition}
    A $\mathcal{A}$-budget-resistant protocol 
    satisfies the following properties: \\
    \textit{\textbf{Sybil-resistance}}: 
    There exists a lower bound on the attack budget of acquiring an expected value of $t$ shares in a Sybil attack, which offers linear scaling of $\Delta d$ with the number of executors independent of $\mathcal{A}$.
    \\
    \textit{\textbf{Bribery-resistance}}: 
    There exists a lower bound on the attack budget of acquiring $t$ shares in a bribery attack, which offers linear scaling of $\Delta d$ with $tl$.
\end{definition}

\section{T-Watch: Protocol}
\label{s5}

In this section, we start by providing an overview of the protocol that implements \textsf{T-Watch}.
We then identify essential requirements and describe a number of key design options.
Finally, we depict the proposed protocol in detail.
Throughout Section~\ref{s5}, we assume that a user is scheduling a function invocation transaction to call a function within a target smart contract denoted as $C_{t}$.
We summarize the notations that will be used in this section in TABLE~\ref{t3}.


\begin{table}
\begin{center}
\begin{tabular}{c p{6.5cm}}
\hline
\textbf{notation} & \textbf{description} \\
\hline
$U_l$ & a user (leader) of Timed Transaction (TT) \\
$U_f$ & a user (follower) of Timed Transaction (TT) \\
$E$ & an executor in TT \\
$\boldsymbol{E}$ & an execurive committee in TT \\
$C$ & a smart contract \\
$C.fun()$ & function $fun()$ within contract $C$ \\
$\rightrightarrows$ & broadcast information via off-chain channels\\
$\dashrightarrow$ & transmit infomation via private off-chain channels \\
$\Rightarrow$ & invoke a function within a contract\\
$addr(*)$ & an address of an EOA or a CA\\
\hline
\end{tabular}
\end{center}
\vspace{-4mm}
\caption{Summary of notations.}      
\vspace{-5mm}
\label{t3}
\end{table}

\subsection{Protocol overview}
\label{s5.1}

In Fig.~\ref{protocol_sketch}, we sketch the overall protocol as a three-phase process for serving \textit{TT} in the context of the committee-driven \textsf{T-Watch} architecture:
(1) \textit{TT.schedule}, a user (leader) establishes a new committee during this phase;
(2) \textit{TT.waiting}, the committee preserves the encrypted \textit{tt.payload} and shares during this phase;
(3) \textit{TT.execute}, the committee executes the timed transaction during this phase.
Next, following the order of these three phases, we provide an overview of the protocol under \textit{T-Opt}, \textit{T-Pes} and \textit{T-Pool}, respectively. 

\noindent \textbf{\textit{T-Opt}}: 
During \textit{TT.schedule}, a user who decides to become a leader ($U_l$) and establish a new committee needs to sequentially perform three operations.
Concretely, the leader first deploys a proxy contract $C_{p}$ with \textit{tt.fund} (if needed) and \textit{tt.timer},
then announces her service requirements and declares the list of a new committee ($\boldsymbol{E}$) at a bulletin-board smart contract denoted as $C_b$,
and finally delivers her encrypted private data to $\boldsymbol{E}$ via off-chain channels. 
After that, no action would be required until \textit{TT.execute}.
During the prescribed time-frame, all the executors within $\boldsymbol{E}$ need to reveal their service private keys $sk_e$ (not their account private keys) via off-chain channels so that they can sequentially decrypt shares, restore $sk_u$, decrypt \textit{tt.payload} and finally execute the timed transaction to call the target contract $C_t$ via $C_{p}$.

\noindent \textbf{\textit{T-Pes}}: 
\textit{T-Opt} expects executors to honestly reveal their service private key $sk_e$ during the prescribed time-frame.
However, in practice, executors may choose to reveal $sk_e$ before the prescribed time-frame, never reveal $sk_e$, or reveal fake $sk_e$.
Despite the use of the $(t,n)$-threshold secret sharing, corrupted $sk_e$ may result in more than $n-t$ unavailable shares.
Therefore, we design \textit{T-Pes} to replace \textit{T-Opt}, 
when \textit{T-Opt} has failed to restore $sk_u$ during the first epoch of \textit{TT.execute}.
Specifically, 
after confirming that \textit{T-Opt} has failed, an executor could deploy a prescribed supplemental contract $C_s$ from the proxy contract $C_p$ and become a watchdog.
Then, executors are incentivized to send real $sk_e$ to $C_s$ within a short time window because any leaked, missing or fake $sk_e$ would be confirmed and penalized after this time window.

\noindent \textbf{\textit{T-Pool}}: 
A follower $U_f$ could join a service request pool associated with $U_l$ and $\boldsymbol{E}$ during \textit{TT.waiting}.
Here, $U_f$ needs to deliver the private data to $\boldsymbol{E}$ via off-chain channels.
The private data of $U_f$ shall be encrypted by the service public key $pk_u$ of $U_l$ who has established $\boldsymbol{E}$ during \textit{TT.schedule}.


   

\subsection{Key design options}

\noindent \textbf{Punishment mechanism}: 
To resolve the problem of lacking a way of pushing executors to behave honestly, we propose to design the protocol with a punishment mechanism that penalizes dishonest executors by confiscating their security deposits and rewards reporters of misbehaviors using confiscated security deposits.
The idea of employing such a punishment mechanism is inspired by recent efforts that leverage cryptocurrency as deposits to improve security, including using bitcoin to penalize anyone who unfairly aborts a secure multiparty computation (SMC)~\cite{andrychowicz2014secure}, as well as using \textit{ether} as security deposits to provide verifiable cloud computing~\cite{dong2017betrayal} or to enforce certificate authorities to be honest~\cite{matsumoto2017ikp}.

\begin{figure}
\centering
{
   
    \includegraphics[width=0.88\columnwidth]{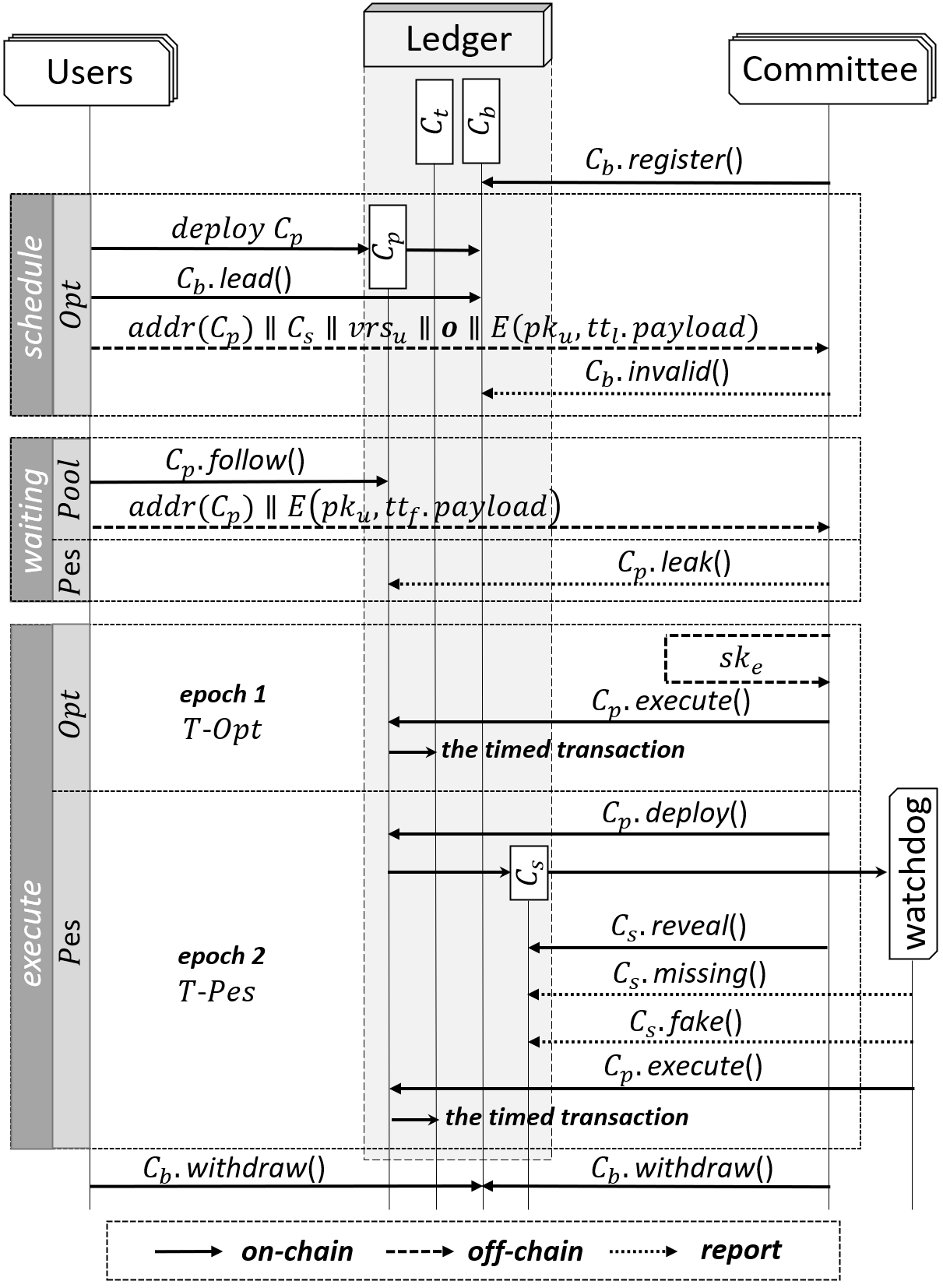}
}
\caption {An overview of the proposed \textsf{T-Watch} protocol.} 
\vspace{-5mm}
\label{protocol_sketch} 
\end{figure}

\noindent \textbf{Reputation-weighted remuneration mechanism}: 
To attract high-quality executors and build a sustainable \textsf{T-Watch} ecosystem, we propose a reputation-weighted remuneration mechanism.
Initially, an executor has a starting reputation score of $r=r_{l}$ and a starting difficulty coefficient of $\tau=1$.
Then, every $\tau$ times this executor successfully completes a \textsf{T-Watch} task, 
the reputation score increases by a step length of $\Delta r$ until it reaches a prescribed upper bound $r=r_{u}$, and the value of $\tau$ also increases by one to enhance the difficulty of improving reputation. 
Meanwhile, after each task, the execution receives remuneration of $r\Delta p$ or $r\Delta p+\zeta$, where $\Delta p$ denotes the amount of remuneration per unit of reputation score, and $\zeta$ denotes a bonus paid to the executor who invokes the execution function at $C_p$, as shown in Fig.~\ref{protocol_sketch}.
However, upon being convicted of any misbehavior, the executor will be immediately blacklisted and lose the score.
The reputation-weighted remuneration mechanism helps further increase the implicit cost of misbehaving, which will be presented with more details in Section~\ref{s6.2}.


\noindent \textbf{Splitting complicated smart contracts}: 
To reduce the gas cost of deploying a complicated smart contract that unnecessarily includes all the functions that support all the three paths, we propose to divide such a complex contract into two separate contracts, namely a proxy contract $C_p$ that always needs to be deployed, as well as a supplemental contract $C_s$ that only needs to be conditionally deployed.
By default, a user deploys $C_p$ during \textit{TT.schedule}, which then serves both \textit{T-Opt} and \textit{T-Pool}.
If needed, an executor deploys $C_s$ to rebind it with $C_p$ during the second epoch of \textit{TT.execute}, which later supports \textit{T-Pes}.


\begin{figure}
\begin{minipage}{0.5\textwidth}
\begin{small}
\begin{mdframed}[innerleftmargin=8pt]

\noindent \textbf{TT.schedule [Opt]:} 
\begin{packed_enum}[leftmargin=*]
  \item 
    A user (leader) $U_l$ creates $C_{p}$ and $C_{s}$ locally, deploys $C_{p}$ with $tt_l.fund$. 
    The deployed $C_{p}$ immediately informs $C_b$ of both the current block number $bn$ and $addr(C_p)$.
    Meanwhile, $U_l$ fetchs the full list of registered executors $\boldsymbol{R}$ at $bn$ from $C_b$.
  \item 
     $U_l$ generates a service keypair $\langle pk_u,sk_u \rangle$, 
     creates a pseudo-random number $r \gets \textit{VH}(sk_u,addr(C_b) \parallel bn)$ and a proof $\pi \gets \textit{VP}(sk_u,addr(C_b) \parallel bn)$.
     Then, $\forall i \in \{1,...,nl\}$, 
     $U_l$ selects $E_i$ for the committee $\boldsymbol{E}$ by picking the first available executor starting from $R_{\textit{H}(r,i) \% |\boldsymbol{R}|}$ in $\boldsymbol{R}$.
     Finally, $U_l \Rightarrow C_{b}.lead(tt_l.timer,pk_u,\{\pi,r,bn\},\{l,t,n\},addr(C_{p}),\boldsymbol{E})$.
\end{packed_enum}

\begin{packed_enum}[leftmargin=*]
  \setcounter{enumi}{2}
  \citem 
    $U_l$ obtains shares $\boldsymbol{s} \gets SS.split(sk_u,\{t,n\})$, 
    and encrypts shares $\boldsymbol{s}$ to onions $\boldsymbol{o}$ using public service keys of $\boldsymbol{E}$, namely
    $\forall i \in \{1,...,|\boldsymbol{E}|\}$, $o_i \gets E(pk_{l(i-1)+l}^e,...,E(pk_{l(i-1)+1}^e,s_i)...)$.
\end{packed_enum}

\begin{packed_enum}[leftmargin=*]
  \setcounter{enumi}{3}
  \citem 
    $\forall i \in \{1,...,|\boldsymbol{E}|\}$,
    $U_l \dashrightarrow E_i$: 
    $addr(C_{p}) \parallel C_s \parallel vrs_u \parallel \boldsymbol{o} \parallel E(pk_u,tt_l.payload)$, 
    where $vrs_u \gets \textit{S}(\textit{H}(addr(C_{p}) \parallel C_s))$.
\end{packed_enum}

\begin{packed_enum}[leftmargin=*]
  \setcounter{enumi}{4}
  \item 
    Upon detecting any invalid $E_j \in \boldsymbol{E}$ uploaded in step 2, any $E_i  \in \boldsymbol{E}$ could report it through 
    $E_i \Rightarrow C_{b}.invalid(j)$.  
\end{packed_enum}

\noindent \textbf{TT.waiting [Pool]:} 
\begin{packed_enum}[leftmargin=*]
  \setcounter{enumi}{5}
  \item 
  Each user (follower) $U_f \Rightarrow C_{p}.follow(tt_f.fund)$.
\end{packed_enum}

\begin{packed_enum}[leftmargin=*]
  \setcounter{enumi}{6}
  \citem 
    $\forall i \in \{1,...,|\boldsymbol{E}|\}$,
    $U_f \dashrightarrow E_i$: 
    $addr(C_{p}) \parallel E(pk_u,tt_f.payload)$.
\end{packed_enum}

\noindent \textbf{TT.waiting [Pes]:} 
\begin{packed_enum}[leftmargin=*]
  \setcounter{enumi}{7}
  \item Upon detecting leakage of any $sk^e_i \in \{sk^e_1,...,sk^e_{nl}\}$, 
  any account $\Rightarrow C_p.leak(sk^e_i,pk^e_i)$.
\end{packed_enum}

\noindent \textbf{TT.execute.epoch-1 [Opt]:}
\begin{packed_enum}[leftmargin=*]
  \setcounter{enumi}{8}
  \citem 
    $\forall i \in \{1,...,|\boldsymbol{E}|\}$,
    $E_i \dashrightarrow \boldsymbol{E}: addr(C_{p}) \parallel sk^e_i$.
  \citem 
    $\forall i$,
    $E_i$ restores $s_i \gets D(sk_{l(i-1)+1}^e,...,D(sk_{l(i-1)+l}^e,o_i)...)$, 
    and the successfully restored $s_i$ forms $\boldsymbol{\hat{s}}$.
  \item 
    If $|\boldsymbol{\hat{s}}| \ge t$, any $E_i$ does the following and the protocol completes with state \texttt{SUCCESS}.
    \begin{packed_enum}[leftmargin=*]
    \item 
      Restore $sk_u \gets \textit{SR}(\boldsymbol{\hat{s}},\{t,n\})$ and obtain 
      $tt_l.payload \gets D(sk_u,E(pk_u,tt_l.payload))$.
    \item 
      $E_i \Rightarrow C_p.execute(tt_l.payload)$.
    \item 
      Repeat step 11.1 and step 11.2 for all $tt_f.payload$ received in step 7.
    \end{packed_enum}
    Otherwise, the protocol \underline{goes to epoch 2 (\textit{T-Pes}).}
\end{packed_enum}

\noindent \textbf{TT.execute.epoch-2 [Pes]:} 
\begin{packed_enum}[leftmargin=*]
  \setcounter{enumi}{11}
  \item 
    Any $E_i \in \boldsymbol{E}$ could deploy $C_{s}$ through $C_{p}$ via $E_i \Rightarrow C_{p}.deploy(addr(C_{p}),C_{s},vrs_u)$, thus turning the protocol from \textit{T-Opt} into \textit{T-Pes}. The executor who deploys $C_{s}$ then becomes a watchdog $W$.
  \item 
    $\forall i \in \{1,...,|\boldsymbol{E}|\}$,
    $E_i \Rightarrow C_{s}$.$\textit{reveal}(sk^e_i)$,
    and the correctly uploaded $s_i$ forms $\boldsymbol{\hat{s}}$.
  \item 
    Upon detecting any missing $sk^e_i$ in step 13, $W$ could report this misbehavior through
    $W \Rightarrow C_{s}.missing(addr(E_i))$. 
  \item 
    Upon detecting any fake $sk^e_i$ in step 13, $W$ could report this misbehavior through
    $W \Rightarrow C_{s}.fake(addr(E_i))$. 
  \item 
    If $|\boldsymbol{\hat{s}}| \ge t$, $W$ does the following and the protocol completes with state \texttt{SUCCESS}.
    \begin{packed_enum}[leftmargin=*]
    \item 
      Restore $sk_u \gets \textit{SR}(\boldsymbol{\hat{s}},\{t,n\})$ and obtain 
      $tt_l.payload \gets D(sk_u,E(pk_u,tt_l.payload))$.
    \item 
      $W \Rightarrow C_p.execute(tt_l.payload)$.
    \item 
      Repeat step 14.1 and step 14.2 for all $tt_f.payload$ received in step 7.
    \end{packed_enum} 
    Otherwise, the protocol completes with state \texttt{FAILURE}.
\end{packed_enum}

\end{mdframed}
\end{small}
\end{minipage}

\captionof{figure}{
The protocol that implements \textsf{T-Watch}. 
A step with a gray bullet (e.g., $\colorbox{gray!28}{3}$) refers to an off-chain operation not recorded by blockchain while a step with a white bullet (e.g., 2) refers to an on-chain operation recorded by blockchain.
}
\vspace{-4mm}
\label{protocol_detail_1}
\end{figure}

\subsection{Protocol detailed description}
\label{s5.3}
We show the detailed protocol in Fig.~\ref{protocol_detail_1}.
Before \textit{TT.schedule}, it is worth noting that the protocol demands each executor to register with the bulletin-board contract $C_b$ through $register()$.
Concretely, an exexutor $E_i$ needs to provide three things to $C_b$:
\begin{itemize}[leftmargin=*]
\item
A public key for facilitating off-chain communications through the Whisper protocol introduced in Section~\ref{s3.4};
\item 
An amount of $k_i\Delta d$ \textit{ether} as a security deposit, where $\Delta d$ denotes the prescribed amount of deposit per service and $k_i$ denotes the maximum number of services $E_i$ intends to simultaneously participate in;
\item 
A total of $m_i$ independent public service keys, each of which is randomly created by $E_i$ in the form of $\langle pk_i^e,sk_i^e \rangle$ for attending a different service, and the corresponding private service keys must be well maintained by $E_i$.
\end{itemize}
Then, every time $E_i$ gets selected for an executive committee to serve a user (leader), one service keypair $\langle pk_i^e,sk_i^e \rangle$ needs to be consumed and deleted.
Consequently, the number of leaders that $E_i$ can simultaneously serve is $min(k_i,\hat{m_i})$, 
namely the smaller one between $k_i$ and $\hat{m_i}$, 
where $\hat{m_i}$ denotes the remaining public service keys recorded in $C_b$.
Besides, $E_i$ can adjust $k_i$ by either transferring more \textit{ether} to $C_b$ or withdrawing unlocked \textit{ether} from $C_b$, and should supplement new public service keys in a timely manner.

\noindent \textbf{\textit{TT.schedule} [Opt]}: 
In the first step, a leader $U_l$ needs to create two contracts $C_p$ and $C_s$ locally.
Specifically, $C_p$ consists of three execution functions and a few utility functions.
The execution functions 
are developed to process the three categories of transactions presented in Section~\ref{s3.2}, respectively, by first verifying the four key components shown in Eq. 2, and then employing EVM-level opcodes \texttt{CALL} and \texttt{CREATE}.
We denote the three execution functions as a single function named \textit{execute()} in the rest of this section.
Besides, a typical utility function is \textit{deploy()}, through which any executor $E_i$ can turn the protocol from \textit{T-Opt} into \textit{T-Pes} by deploying $C_{s}$ that contains additional utility functions such as \textit{reveal()}, \textit{leak()}, \textit{missing()} and \textit{fake()}. 
After creating the two contracts, $U_l$ needs to deploy $C_p$, 
which immediately notifies both the current block number $bn$ and the address $addr(C_p)$ to $C_b$.
Meanwhile, $U_l$ needs to fetch the full list of registered executors $\boldsymbol{R}$ and their status of availability at $bn$ from $C_b$.


In step 2, 
$U_l$ needs to first create an independent service keypair $\langle pk_u,sk_u \rangle$, where $pk_u$ would later be used by both $U_l$ and potential followers $U_f$ to encrypt their private data sent to $\boldsymbol{E}$. 
After that, $U_l$ could obtain a verifiable pseudo-random number $r \gets \textit{VH}(sk_u,addr(C_b) \parallel bn)$ and a corresponding proof $\pi \gets \textit{VP}(sk_u,addr(C_b) \parallel bn)$, where $addr(C_b) \parallel bn$ is used as the input message to make $r$ unique to $U_l$ at $bn$.
To construct an executive committee $\boldsymbol{E}$, $U_l$ needs to select the $i$th member $E_i$ of $\boldsymbol{E}$ by pseudo-randomly finding a registered executor that has its index in $\boldsymbol{R}$ to be $\textit{H}(r,i) \% |\boldsymbol{R}|$ and then picking the first available executor starting from $R_{\textit{H}(r,i) \% |\boldsymbol{R}|}$ in $\boldsymbol{R}$ as $E_i$.
In this way, given $r$ and $bn$, $\boldsymbol{E}$ is deterministic and thus verifiable.
Finally, $U_l$ needs to inform $C_b$ about the detailed information, including an expected time-frame $tt_l.timer$, the public service key $pk_u$, $\{\pi,r,bn\}$ required for proving $\boldsymbol{E}$, three parameters $\{l,t,n\}$, $addr(C_p)$ and $\boldsymbol{E}$, and finally an amount of \textit{ether} consisting of a security deposit to store in $C_b$ and a remuneration to pay $\boldsymbol{E}$.

After that, in step 3, $U_l$ needs to further split $sk_u$ into $n$ shares $\boldsymbol{s}=\{s_1,...,s_n\}$ based on $(t,n)$-threshold secret sharing.
Instead of directly sending shares $\boldsymbol{s}$ to executors $\boldsymbol{E}$, each share $s_i$ needs to be iteratively encrypted with $l$ public service keys belonging to different executors, which thus makes the size of $\boldsymbol{E}$ $nl$, as presented in step 2.
In this way, each share $s_i$ could be turned into an onion $o_i$ and its recovery needs the corresponding $l$ private service keys maintained by the same set of executors.
This design helps turn the leakage of a share $s_i$ into the leakage of $l$ private service keys so that the corresponding executors could be held accountable because each leaked $sk^e_i$ could be verified using a $pk^e_i$ stored in $C_b$ and $sk^e_i$ is known only by $E_i$.
We will later discuss how this design can help make the protocol resilient against Sybil attacks in Section~\ref{s6}.

Finally, in step 4, through private off-chain channels, $U_l$ transmits $tt_l.payload$ encrypted by $pk_u$ to $\boldsymbol{E}$, along with $addr(C_{p}) \parallel C_s$ and a corresponding signature $vrs_u$, as well as all the onions $\boldsymbol{o}$.
In step 5, given $pk_u$ and $\{\pi,r,bn\}$, any $E_i$ could first verify $r$ through $r \gets \textit{VV}(pk_u,addr(C_b) \parallel bn,\pi)$ and then verify $\boldsymbol{E}$ using $r$ and $bn$. If either $r$ or $\boldsymbol{E}$ is invalid, $E_i$ could call \textit{invalid()} to make $C_b$ verify $r$ and a chosen $E_j$ in $\boldsymbol{E}$.
In case either $r$ or $E_j$ is proved to be invalid by $C_b$, the security deposit of $U_l$ would be transferred to $E_i$, the service would be canceled, and $U_l$ would be blacklisted.

\noindent \textbf{\textit{TT.waiting} [Pool]}: 
During \textit{TT.waiting}, a follower $U_f$
could reuse the already deployed $C_p$ and $\boldsymbol{E}$ to schedule a transaction to be executed during the same time-frame prescribed by $U_l$.
Specifically, in step 6, $U_f$ needs to first inform $C_p$ about a new follower by calling a utility function named \textit{follow()} and sending $tt_f.payload$ to $C_p$.
Besides, $U_f$ also needs to transfer a small amount of \textit{ether} to $C_p$ to pay the executor who will later trigger $C_p$ to serve $U_f$, and also the leader $U_l$ for the efforts of establishing $C_p$ and $\boldsymbol{E}$.
After that, in step 7, $U_f$ needs to encrypt $tt_l.payload$ using $pk_u$ and transmit the encrypted payload data to $\boldsymbol{E}$. 
It is worth noting that, by default, no restrictions are placed upon the maximum number of followers of a leader, but it is easy for leaders to set a restriction through $C_p$.

\noindent \textbf{\textit{TT.waiting} [Pes]}:
Meanwhile, during \textit{TT.waiting}, the protocol stipulates that executors need to protect their private service keys well, but it is possible that some executors disclose their keys to earn profit.
Due to the various online or offline ways of disclosing a $sk^e_i$, it is difficult to proactively detect leakage, but we could incentive anyone who deliberately or accidentally finds out a $sk^e_i$ to report it to $C_p$.
Thus, step 8 indicates that, during \textit{TT.waiting}, any account in Ethereum could report a leaked $sk^e_i$ by calling a reporting function named \textit{leak()} and indicating the corresponding $pk^e_i$.
If the leakage is confirmed by $C_p$, the security deposit of the executor that owns the leaked $sk^e_i$ will be confiscated and further split into two parts, a larger part that rewards the reporter, as well as a smaller part that rewards $U_l$.
The protocol splits the reward to discourage executors from intentionally reporting themselves to withdraw their security deposits.

\noindent \textbf{\textit{TT.execute.epoch-1} [Opt]}:
The \textit{TT.execute} phase consists of two epochs, which corresponds to \textit{T-Opt} and \textit{T-Pes}, respectively.
At \textit{epoch-1}, namely \textit{T-Opt}, executors are required to reveal their private service keys to each other via off-chain channels (step 9), and then try to decrypt and restore as many shares as possible (step 10).
After that, in step 11, depending on the number of successfully restored shares, the protocol branches out into two directions.
If there are more than $t$ available shares, executors will be able to recover $sk_u$, decrypt both $tt_l.payload$ and $tt_f,payload$, and call the execution function \textit{execute()}, which will make the protocol complete with state \texttt{SUCCESS}.
Otherwise, the protocol could not come to state \texttt{SUCCESS} at the end of \textit{TT.execute}, and it will have to enter \textit{epoch-2}.

\noindent \textbf{\textit{TT.execute.epoch-2} [Pes]}:
In step 12, any executor could deploy $C_s$ by invoking the utility function \textit{deploy()} at $C_p$, which turns the protocol from \textit{T-Opt} into \textit{T-Pes}.
The deployer of $C_s$ automatically becomes a watchdog denoted as $W$, whose invocations of execution or reporting functions in the remaining steps will be prioritized, while invocations made by other executors must wait for a short period of time assigned to $W$.
Then, in step 13, within a short time window, executors in $\boldsymbol{E}$ need to reveal their private service keys to $C_s$ using the function \textit{reveal()}.
If there is any missing or fake $sk_e$, $W$ could report it by calling the reporting functions \textit{missing()} and \textit{fake()}, in step 14 and 15, respectively.
Finally, similar to step 10, depending on the number of available shares, the protocol branches out into two final states in step 16, namely \texttt{SUCCESS} and \texttt{FAILURE}.
Besides, the protocol offers a function named \textit{withdraw()} that allows participants to withdraw their unlocked \textit{ether} from $C_b$ at any time.

\section{Security analysis}
\label{s6}

In this section, we analyze the security of the proposed \textsf{T-Watch} protocol regarding the two threat models, $\mathcal{A}$-threshold and $\mathcal{A}$-budget, introduced in Section~\ref{s4.3}.

\subsection{Security against $\mathcal{A}$-threshold}
\label{s6.1}

\begin{theorem}
    \label{theorem:1}
    Protocol \textsf{T-Watch} is $\mathcal{A}$-threshold-resistant by Definition 1.
\end{theorem}

Recall that Definition 1 requires a secure protocol to satisfy the secrecy property. We prove it in Lemma~1.

\begin{lemma}
    Protocol \textsf{T-Watch} satisfies secrecy.
\end{lemma}

\begin{proof}
    As presented in Fig.~\ref{protocol_detail_1}, the protocol contains eight steps before \textit{TT.execute}.
    Besides the registration information of executors, the information acquired by the adversary $\mathcal{A}$, who is able to corrupt up to $tl-1$ executors in the executive committee $\boldsymbol{E}$, includes:
    \begin{itemize}[leftmargin=*]
      \item Before step~1:
        for each $E_i$, $addr(E_i)$ and $pk^e_i$; 
        in case of a corrupted $E_i$, $sk^e_i$ as well;
      \item Step~1: 
        $addr(U_l)$, $addr(C_{p})$, $C_{p}$, $tt_l.fund$;
      \item Step~2: 
        $pk_u$, $\{\pi,r,bn\}$, $\{l,t,n\}$, $\boldsymbol{E}$, $tt_l.timer$;
      \item Step~4: 
        $C_s$, $E(pk_u,tt_l.payload)$, $vrs_u$, $\boldsymbol{o}$;
      \item Step~6: 
        for each $U_f$, $addr(U_f)$ and $tt_f.fund$;
      \item Step~7: 
        for each $U_f$, $E(pk_u,tt_f.payload)$.
    \end{itemize}
    The information above suggests that $\mathcal{A}$ could learn 
    $pk^e_i$ for each $E_i$ before step 1, 
    $pk_u$, $\{\pi,r,bn\}$ and $\{l,t,n\}$ in step 2, 
    $E(pk_u,tt_l.payload)$, $vrs_u$ and $\boldsymbol{o}$ in step 4, 
    and $E(pk_u,tt_f.payload)$ in step 6.
    Besides, in the worst case, $\mathcal{A}$ corrupts $tl-1$ executors in $\boldsymbol{E}$ and obtains their private service keys. 
    The remaining information is obviously independent of $sk_u$.
    Next, we complete the proof as follows:
    \begin{itemize}[leftmargin=*]
      \item  
          First, based on the assumption of the hardness of the Elliptic Curve Discrete Logarithm problem (ECDLP) underpinning ECDSA, $\mathcal{A}$ learns no extra information about $sk_u$ from $pk_u$, $E(pk_u,tt_l.payload)$ and $E(pk_u,tt_f.payload)$, and learns no extra information about $sk_e$ from $pk_e$ and $\boldsymbol{o}$. 
      \item 
          Then, based on the assumption of the hardness of the Decisional Diffie-Hellman (DDH) problem  underpinning ECVRF, $\mathcal{A}$ learns no extra information about $sk_u$ from $\{\pi,r,bn\}$. 
      \item 
          Finally, given $\{l,t,n\}$ and $tl-1$ executors' private service keys, in the worst case, $\mathcal{A}$ could obtain $t-1$ shares of $sk_u$.
          Due to the information theoretic security of Shamir's secret sharing scheme,  $\mathcal{A}$ learns no extra information about $sk_u$ from the obtained $t-1$ shares. 
    \end{itemize}
    Therefore, protocol \textsf{T-Watch} satisfies secrecy.
\end{proof}

\subsection{Security against $\mathcal{A}$-budget}
\label{s6.2}

\begin{theorem}
    \label{theorem:2}
    Protocol \textsf{T-Watch} is $\mathcal{A}$-budget-resistant by Definition 2.
\end{theorem}

Recall that Definition 2 requires a secure protocol to satisfy Sybil-resistance and bribery-resistance. We prove the two properties sequentially.

\begin{lemma}
    Protocol \textsf{T-Watch} satisfies Sybil-resistance.
\end{lemma}

\begin{proof}
    The adoption of ECVRF ensures both uniqueness and pseudorandomness of $r$, which is then used in constructing $\boldsymbol{E}$.
    Next, we complete the proof as follows:
    \begin{itemize}[leftmargin=*]
      \item  
           We divide all the registered executors (not just the ones in $\boldsymbol{E}$) into two groups, a group of $g_s$ executors controlled by $\mathcal{A}$, and a group of $g_o$ executors that are independent of $\mathcal{A}$. 
      \item 
          Then, to acquire a certain share, $\mathcal{A}$ needs to control the $l$ corresponding executors all together to get their private service keys, which gives a probability of $p=(\frac{g_s}{g_s+g_o})^l$.
      \item 
          Given a total of $n$ shares, by modeling the problem as a binomial distribution with $n$ and $p$, the attack budget of acquiring an expected value of $t$ shares is $b=\frac{t g_s \Delta d}{np}$.
      \item
          To compute the lower bound of $b$, we make $\frac{\partial b}{\partial g_s}=0$:
          \begin{align*}
            l (g_s+g_o)^{l-1} x^{l-1} &= (l-1) (g_s+g_o)^{l} x^{l-2} \\
                                l g_s &= (l-1) (g_s+g_o) \\
                                  g_s &= (l-1) g_o 
          \end{align*}
    \end{itemize}
    Thus, $b$ is bounded by $(l-1) g_o \Delta d$, which proves that protocol \textsf{T-Watch} satisfies Sybil-resistance.
\end{proof}

\begin{lemma}
    Protocol \textsf{T-Watch} satisfies bribery-resistance.
\end{lemma}

\begin{proof}
    Due to the mutual distrust between $\mathcal{A}$ and an independent executor, during bribery, the executor would assume that $\mathcal{A}$ will report the leakage to $C_p$, which will potentially make the executor get blacklisted and lose both reputation score and security deposit. 
    Thus, to bribe a rational executor, $\mathcal{A}$ has to pay the executor an amount of \textit{ether} higher than the executor's potential loss, which consists of two parts:
    \begin{itemize}[leftmargin=*]
      \item  
          Punishment mechanism: an amount of $\Delta d$ due to the loss of security deposit;
      \item 
          Reputation-weighted remuneration mechanism: an amount of 
          $\sum_{i=0}^{\frac{r-r_l}{\Delta r}}  i (\frac{r-r_l}{\Delta r}-i+1) \Delta r \Delta p$ 
          due to the lose of remuneration during the reconstruction of a reputation score of $r$ from the starting score of $r_l$ by registering a new executor.
    \end{itemize}
     Thus, $b$ is bounded by 
     $\sum_{j=0}^{tl} (\Delta d + \sum_{i=0}^{\frac{r_j-r_l}{\Delta r}}  i (\frac{r_j-r_l}{\Delta r}-i+1) \Delta r \Delta p)$
     , which proves that protocol \textsf{T-Watch} satisfies bribery-resistance.
\end{proof}

\noindent \textbf{Remark}: 
From the analysis above, we could observe that the security deposit $\Delta d$ introduced in the punishment mechanism plays a key role in the budget $b$ in both Sybil and bribery attacks.
Besides, we could see that $b$ in a Sybil attack is independent of both the threshold $t$ and the size $g_s$ of the group of executors controlled by $\mathcal{A}$, but tends to be proportional to $l$ and $g_o$, which suggests that users may dynamically adjust $l$ to customize the resistance of their services against Sybil attacks, and may choose to set a relatively smaller $l$ when the scale of registered executors is large.
Finally, we could see that the reputation-weighted remuneration mechanism helps further increase $b$ because of the additional compensation paid to betrayers, especially those with high reputation scores.

\subsection{Discussion}

\noindent \textbf{Security against rational or malicious users}:
Recall that, to discourage executors from intentionally reporting themselves, the protocol splits confiscated security deposits into two parts and rewards both reporters and users.
This design, however, may incentivize rational or malicious users to cooperate with $\mathcal{A}$ to defraud honest executors.
In the worst case, a user (leader) may choose to become $\mathcal{A}$.
Thanks to the careful design of onions, our protocol intentionally prevents any executor from obtaining the plaintext of any share because shares are also known by leaders.
Instead, executors would be panelized only when their private service keys are leaked, missing or fake.
In addition, executors are required to create independent keypairs for different services, so keys for a certain service would be random and independent of keys for other services.
Consequently, based on the assumption of the hardness of ECDLP, our protocol can protect honest executors from getting defrauded by rational or malicious users.

\noindent \textbf{Security of followers' $tt$}:
Recall that in the \textit{T-Pool} path, a follower $U_f$ needs to rely on an existing committee $\boldsymbol{E}$ established by a leader $U_l$. 
Therefore, the security of followers' $tt$ relies on the pseudorandomness of $\boldsymbol{E}$.
In the worst case that $U_l$ is $\mathcal{A}$,
due to the use of ECVRF, the pseudorandomness of $r$ could be guaranteed.
Then, based on the proof of Lemma 2, $b$ in a Sybil attack is independent of the threshold $t$, which suggests that the strategy of repeatly recomputing $r$ to make more controlled executors get involved in $\boldsymbol{E}$ is not helpful.
In addition, the payments made by followers to leaders in step 5 of the proposed protocol would incentivize leaders to select executors in a more transparent and trustworthy manner so that they could attract more followers and earn more profit.
Besides, as long as there is a single rational executor in $\boldsymbol{E}$, followers' timed transactions would be executed as expected.

\noindent \textbf{Rationality of participants}:
\textcolor{black}{The underlying assumption in our threat model $\mathcal{A}$-budget, similar to recent works~\cite{dong2017betrayal,matsumoto2017ikp,ge2022shaduf,wadhwa2022he} , is that participants are rational. This assumption is prevalent in current blockchain and cryptographic research and posits that participants will avoid actions that result in a net loss, such as forfeiting a security deposit. While this approach is effective in mitigating most forms of misbehavior among rational actors, it is acknowledged that it may not deter all types of misbehavior, especially those conducted by irrational or extremely adversarial participants.
As such, our future work will focus on enhancing the robustness of our mechanisms to include strategies that can handle scenarios involving irrational or non-economic driven misbehaviors.} 

\noindent \textbf{Switching mechanism from \textit{T-Opt} to \textit{T-Pes}}:
\textcolor{black}{The switching mechanism activates only under strict conditions, specifically for detected misbehavior, ensuring it is used when necessary. Typically, \textit{T-Opt} operates as the default mode, focusing on efficiency with low overhead unless misbehavior is detected, which triggers a switch to the resource-intensive \textit{T-Pes}. The executor initiating this switch, typically the reporter of the misbehavior, has their costs covered by the security deposit of the offending party, ensuring honest participants don't bear the cost. 
Additionally, the system discourages unnecessary \textit{T-Pes} switches. Rational executors won’t switch without verified misbehavior, maintaining operational integrity and cost-effectiveness, as they receive no compensation without a genuine breach. This setup keeps executors incentivized to stay in \textit{T-Opt}, optimizing system efficiency and reducing unnecessary overhead.}

\noindent \textbf{Allocation proportion of deposits}:
\textcolor{black}{The allocation of security deposits involves a key trade-off: encouraging misbehavior reporting and preventing dishonest executors from self-reporting for gain. Our protocol first covers the cost executors face when switching from \textit{T-Opt} to \textit{T-Pes}.
After this, the remaining deposit is equally divided between reporters and users, fostering a balanced incentive structure that promotes vigilant reporting without giving undue advantage to any party. Furthermore, our protocol accommodates varying security needs and risk tolerances across different applications and users by allowing customization of this allocation proportion. This flexibility ensures the protocol's effective integration into diverse environments, meeting specific security demands.}

\section{Implementation and evaluation}

\begin{table}
\begin{center}
\begin{tabular}{cccc}
\hline
    \textbf{Step} & \textbf{Function} & \textbf{Gas consumption} & \textbf{Cost in UDS (\$)} \\ \hline
    1 & deploy $C_{p}$ & 1114612 & 5.08 \\ 
    2 & \textit{lead()} & 797432 & 3.64 \\ 
    5 & \textit{invalid()} & 2196769 & 10.02 \\ 
    6 & \textit{follow()} & 31198 & 0.14 \\ 
    8 & \textit{leak()} & 1264782 & 5.78 \\ 
    11/16 & \textit{execute()} & 108542 & 0.49 \\ 
    12 & \textit{deploy()} & 2419116 & 11.04 \\
    13 & \textit{reveal()} & 89727 & 0.41 \\
    14 & \textit{missing()} & 65766 & 0.30 \\
    15 & \textit{fake()} & 1279726 & 5.85 \\ \hline
\end{tabular}
\end{center}
\vspace{-3mm}
\caption{Gas consumption of key functions.}      
\vspace{-5mm}
\label{t4}
\end{table}

In this section, we implement and evaluate the proposed \textsf{T-Watch} architecture and protocol.
The implementation is designed to run on \textit{rinkeby}~\cite{Rinkeby2017}, the Ethereum official test network where researchers and developers can acquire free \textit{ether} to afford gas consumption of testing protocols and smart contracts.
\textcolor{black}{It is widely recognized and supported by recent studies\cite{ekparinya2019attack,chen2022blockchain,saad2021syncattack} that using test networks, which are designed to closely emulate real-world production environments, is an essential and accepted practice for the preliminary validation of blockchain technologies, providing a safe and cost-effective setting for systematic problem identification and solution refinement.}
Similar to recent work~\cite{dziembowski2019perun,das2018yoda}, we first measure gas consumption of key functions in the proposed protocol because the monetary cost of paying gas fee is unavoidable in running services over public blockchains like Ethereum.
Then, as introduced in Section~\ref{s3.3}, the gas system places greater demand on the scalability of \textsf{T-Watch}, so we vary the size of the executive committee and measure the corresponding cost of running \textsf{T-Watch} through the three execution paths introduced in Section~\ref{s4.2}, respectively.
After that, we compare \textsf{T-Watch} with two related works introduced in Section~\ref{s2}, namely \textsf{TimedExe}~\cite{li2018decentralized-2} for timed execution of function invocation transaction, and \textsf{Kimono}~\cite{Kimono} for timed release of private data.
Finally, \textsf{T-Watch} novelly divides users into two categories, leaders and followers, to further reduce service costs by employing service request pooling. 
To figure out the effectiveness of this design, we vary the number of followers' requests in a service request pool and measure the corresponding expenses afforded by leaders and followers, respectively.
Besides, it is worth noting that there exists an upper bound for the sum of gas usage of all the transactions within a single block and hence, the evaluation of time overhead of executing functions in Ethereum, which is usually on the scale of hundreds of milliseconds, is omitted.

\vspace{-3mm}
\subsection{Gas consumption of key functions}
In TABLE~\ref{t4}, we list the key functions in the order that they appear in the protocol shown in Fig.~\ref{protocol_detail_1}.
For each key function, we measure its gas consumption, which is then converted into USD for ease of understanding.
Concretely, given gas usage $c_g$, the corresponding cost in USD, namely $c_u$, could be computed using 
$c_u=c_g p_g p_e$, where $p_g$ refers to gas price and $p_e$ denotes \textit{ether} price.
Since both $p_g$ and $p_e$ fluctuate wildly, we collects their historical values from \textit{Etherscan}~\cite{etherscan}, and computes their median values over a six-year period from 8/1/2015 to 7/31/2021.
According to the results, in the rest of this section, we set $p_g$ and $p_e$ to be $2.29 \times 10^{-8}$ ether/gas and 199.73 USD/ether, respectively.
Besides, we assume a committee formed by 30 executors in this experiment, as well as an execution function \textit{execute()} that will call an existing smart contract and change the value of a single variable there in the rest of this section.
As can be seen from the results in TABLE~\ref{t4}, 
there are two types of operations that are more gas-consuming, namely deploying smart contracts and verifying private service keys or random numbers.
Specifically, deploying $C_p$ in step 1 costs \$5.08, and deploying $C_s$ through \textit{deploy()} in step 12 spends \$11.04, which demonstrates the need for splitting a single complex contract into the two.
We could also see that verifying a leaked or fake $sk_i^e$ via \textit{leak()} or \textit{fake()} spends around \$5.8, and verifying a random number $r$ via \textit{invalid()} spends \$10.02, which indicates the expensiveness of verifying ECDSA keys or ECVRF random numbers on-chain.
Among the five remaining functions, \textit{lead()} is the most gas-consuming one because it stores a list of recruited executors in $C_b$.
In contrast, \textit{follow()} is very cheap because it only stores a single address in $C_p$.

\begin{figure}
  \centering
  {
     
      \includegraphics[width=0.83\columnwidth]{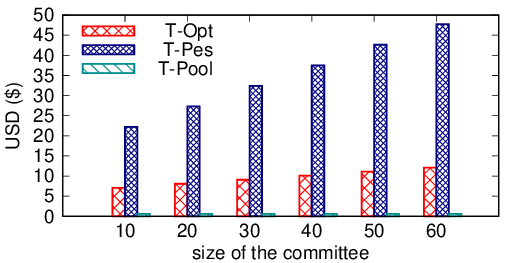}
  }
  \vspace{-2mm}
  \caption {Cost of \textit{T-Opt}, \textit{T-Pes} and \textit{T-Pool}}
  \vspace{-5mm}
  \label{r_committee} 
  \end{figure}

\subsection{Scalability of different execution paths}
In this set of experiments, we first match execution paths with functions and then analyze their scalability.
First, \textit{T-Opt} requires a leader to deploy $C_p$ (\$5.08) and call \textit{lead()} (\$3.64) during \textit{TT.schedule} to set up a new service request, and later an executor to call \textit{execute()} (\$0.49) during \textit{TT.execute} to complete the service, 
which results in a total cost of about \$9.21 assuming 30 executors.
Besides the \$9.21 above, \textit{T-Pes} additionally requires an executor to invoke \textit{deploy()} (\$11.04) to turn the protocol from \textit{T-Opt} to \textit{T-Pes} and later all the $nl$ executors in $\boldsymbol{E}$ to reveal their private service keys via \textit{reveal()} (\$0.41$nl$), which increases the total cost from \$9.21 to around \$32.55.
Different from these two execution paths, \textit{T-Pool} only requires a follower to invoke a single function \textit{follow()} (\$0.14) during \textit{TT.waiting} and hence, its total cost is just \$0.14.
Next, to evaluate the scalability of the execution paths, we measure their overall cost by varying the size of the committee from 10 to 60.
In Fig.~\ref{r_committee}, we can see that the results demonstrate our analysis presented in Section~\ref{s4.2}.
Concretely, during \textit{TT.schedule}, the cost of uploading addresses of $nl$ executors in \textit{lead()} leads to an increase of about \$0.1 per executor in the overall cost of \textit{T-Opt}.
Besides, during \textit{TT.execute}, \textit{T-Pes} needs to afford the additional expenses for uploading $nl$ private service keys via \textit{reveal()}, which further increases the overall cost by around \$0.49 per executor and make \textit{T-Pes} less scalable than \textit{T-Opt}.
Finally, we could see that the overall cost of \textit{T-Pool} stabilizes at \$0.14, making \textit{T-Pool} the most scalable execution path among the three.

\begin{figure}
  \centering
  {
     
      \includegraphics[width=0.83\columnwidth]{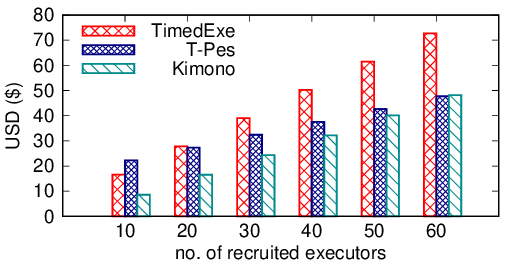}
  }
  \vspace{-2mm}
  \caption {Comparison against \textsf{TimedExe} and \textsf{Kimono}}
  \vspace{-5mm}
  \label{r_comparison} 
\end{figure}

\vspace{-1mm}
\subsection{Comparison against \textsf{TimedExe} and \textsf{Kimono}}
In Fig.~\ref{r_comparison}, we show the overall cost of \textit{T-Pes}, the least scalable execution path of \textsf{T-Watch}, and that of two related works, \textsf{TimedExe} and \textsf{Kimono}, by varying the number of recruited executors from 10 to 60.
We could see that initially, \textit{T-Pes} has the highest cost because the supplemental contract $C_s$ that costs $\$11.04$ to deploy contains some preventive functions such as \textit{missing()} and \textit{fake()}, as well as some utility functions that help associate $C_s$ with $C_p$.
Then, as the number of executors increases, we could see that the cost of \textsf{TimedExe} and \textsf{Kimono} rises above that of \textit{T-Pes} at the moment of 20 and 60 executors, respectively.
The reason is that, \textit{T-Pes} is designed to discourage potential misbehaviors by leveraging the punishment mechanism, while \textsf{TimedExe} and \textsf{Kimono} need to store more data per executor or perform more complicated calculations (e.g., generating shares and recovering secrets) on the blockchain.
Finally, by further comparing \textit{T-Opt} and \textit{T-Pool} with the two related works, we could see that \textit{T-Opt} and \textit{T-Pool} reduces the cost of recruiting 30 executors by over 62\% and 97\%, respectively, and a service request pool of size 4 reduces the cost by over 90\%.

\vspace{-1mm}
\subsection{Effectiveness of service request pooling}
\begin{figure}
\centering
{
   
    \includegraphics[width=0.83\columnwidth]{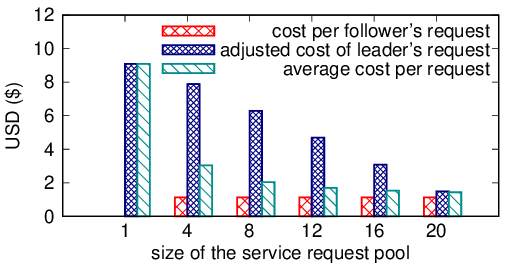}
}
\vspace{-2mm}
\caption {Effectiveness of service request pooling}
\vspace{-5mm}
\label{r_batch}
\end{figure}

In the last set of experiments, we create service request pools of users of different sizes and evaluate the change in expenses.
As shown in Fig.~\ref{r_batch}, each service request pool consists of a single leader and a varying number of followers, \textcolor{black}{adhering to the protocol described in Section~\ref{s4.2}. This experimental setup validates the practicality of our cost-reduction strategy, as delineated in the \textit{T-Pool} path.}
Besides, we make the following assumptions:
(1) leaders' and followers' requests follow \textit{T-Opt} and \textit{T-Pool}, respectively;
(2) the size of the committee is 30;
(3) each follower pays \$0.4 worth of \textit{ether} along with \textit{follow()} to offset leader's expenses, as required by the protocol. 
As can be seen from the results, the cost per follower's request is constant and independent of the service request pool size.
After compensation, the adjusted cost of the leader's request decreases as the service request pool size increases and could reach about \$1.49 when there are 19 followers.
Finally, we could see that the average cost per request in a service request pool decreases rapidly with the rise in service request pool size.
It reduces to about \$3, \$2 and \$1.4 when the service request pool size is 4, 8 and 20, respectively.
The results demonstrate the effectiveness of service request pooling and suggest that even a small service request pool can significantly reduce the average cost per request.

\vspace{-1mm}
\subsection{Off-chain communication cost}
\textcolor{black}{
As demonstrated in Fig. 4 and 5, off-chain communication in the T-Watch protocol primarily occurs in three steps:}

\begin{itemize}[leftmargin=*]
    \item Step~4, part of the \textit{T-Ops} phase, where the leader $U_l$ sends data to each executor $E_i$ in the committee $\boldsymbol{E}$:
    $addr(C_{p}) \parallel C_s \parallel vrs_u \parallel \boldsymbol{o} \parallel E(pk_u,tt_l.payload)$;
    \item Step~7, during the \textit{T-Pool} phase, where each follower $U_f$ sends their transaction payload to each executor $E_i$:
    $addr(C_{p}) \parallel E(pk_u,tt_f.payload)$;
    \item Step~9, part of the \textit{T-Ops} phase, where each executor $E_i$ shares their private service key with the rest of the committee $\boldsymbol{E}$:
    $addr(C_{p}) \parallel sk^e_i$.
  \end{itemize}

\textcolor{black}{To optimize off-chain communication costs, we have employed the InterPlanetary File System (IPFS)~\cite{benet2014ipfs} to efficiently transmit data through off-chain channels. Specifically, instead of transmitting complete data such as $addr(C_{p}) \parallel sk^e_i$, we transmit only the hash values associated with the IPFS stored data, e.g., $hash(addr(C_{p}) \parallel sk^e_i)$. This approach significantly reduces the volume of off-chain communication while ensuring data integrity and immutability.}

\textcolor{black}{Consequently, with $k$ executors and $f$ followers, the off-chain communication cost for \textit{T-Opt} is calculated as $32k + 32k(k-1) = 32k^2$ bytes. For the \textit{T-Pool} phase, specifically in step 7, the off-chain communication cost is $32fk$ bytes. }

\section{Conclusion}
This paper proposes \textsf{T-Watch}, the first practical decentralized solution for cost-effectively implementing timed transactions with strong security and scalability guarantees. 
Our solution employs a novel combination of threshold secret sharing and decentralized smart contracts.
To protect the private elements of a scheduled transaction from getting disclosed before the future time-frame, \textsf{T-Watch} maintains shares of the decryption key of the scheduled transaction using a group of executors recruited in a blockchain network before the specified future time-frame and restores the scheduled transaction at a proxy smart contract to trigger the change of blockchain state at the required time-frame.
To reduce the cost of smart contracts execution in \textsf{T-Watch}, we carefully design a protocol that offers three execution paths with different characteristics.
We rigorously analyze the security of \textsf{T-Watch} and proved that \textsf{T-Watch} is both $\mathcal{A}$-threshold-resistant and $\mathcal{A}$-budget-resistant.
Finally, we implement \textsf{T-Watch} over the Ethereum official test network and the results demonstrate that \textsf{T-Watch} is more scalable compared to the state of the art and could reduce the cost of running smart contracts by over 90\% through pooling.

\renewcommand\refname{Reference}
\bibliographystyle{plain}
\urlstyle{same}

\bibliography{main.bib}

\begin{IEEEbiography} [{\includegraphics[width=1in,height=1.25in]{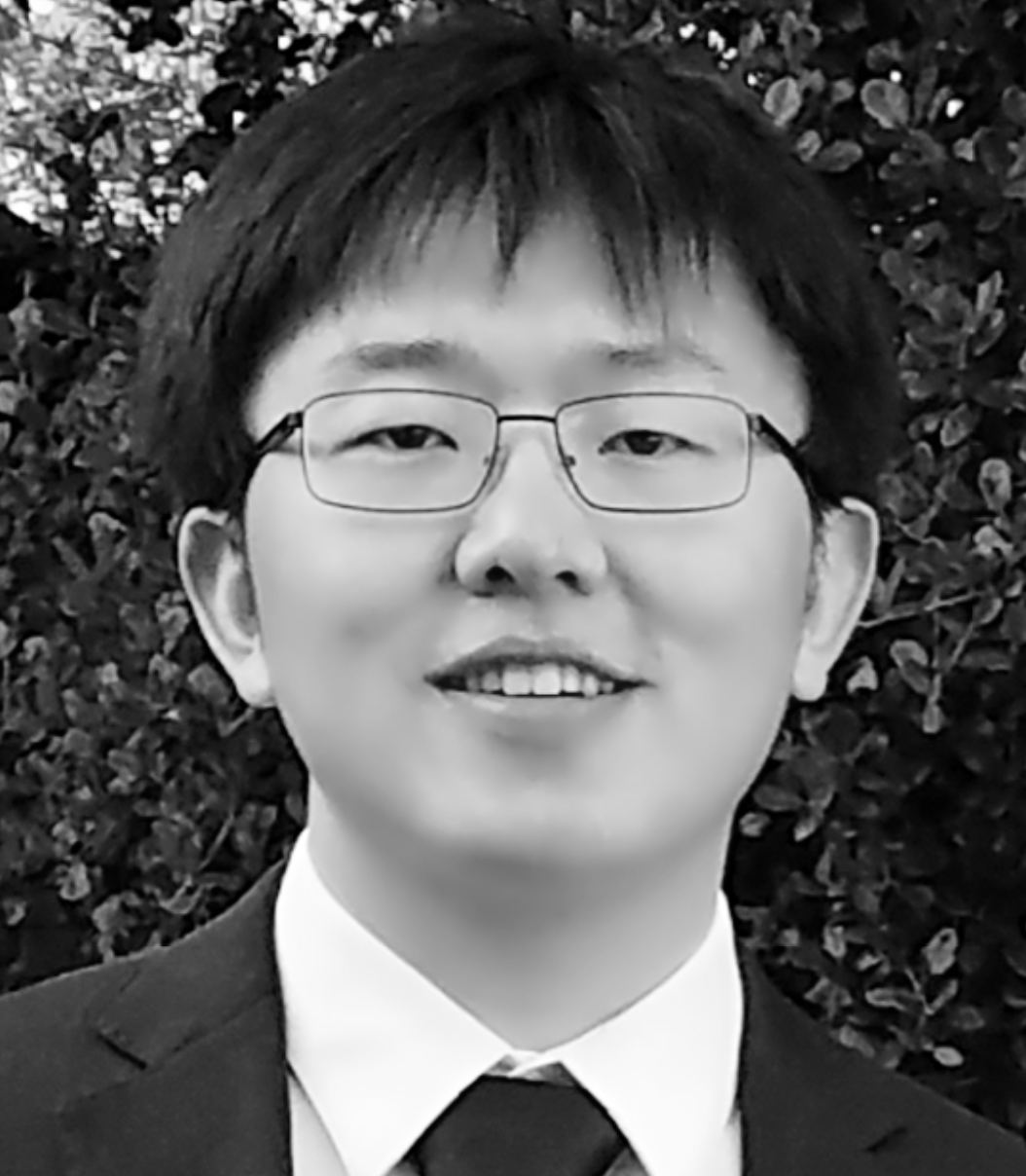}}]{Chao Li}
  is an Associate Professor in the School of Computer and Information Technology at Beijing Jiaotong University. He received his Ph.D. degree from the School of Computing and Information at University of Pittsburgh and his MSc degree from Imperial College London. His current research interests are focused on Blockchain and Data Privacy. 
\end{IEEEbiography}

\vskip 0pt plus -1fil

\begin{IEEEbiography}
  [{\includegraphics[width=1in,height=1.25in]{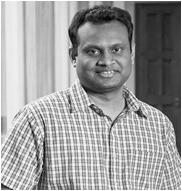}}]{Balaji Palanisamy}
is an Associate Professor in the School of computing and information in University of Pittsburgh. He received his M.S and Ph.D. degrees in Computer Science from the college of Computing at Georgia Tech in 2009 and 2013, respectively. His primary research interests lie in scalable and privacy-conscious resource management for large-scale Distributed and Mobile Systems. 
\end{IEEEbiography}

\end{document}

%% file: solidity-highlighting.tex
\usepackage{listings, xcolor}

\definecolor{verylightgray}{rgb}{.97,.97,.97}

\lstdefinelanguage{Solidity}{
	keywords=[1]{anonymous, assembly, assert, balance, break, call, callcode, case, catch, class, constant, continue, constructor, contract, debugger, default, delegatecall, delete, do, else, emit, event, experimental, export, external, false, finally, for, function, gas, if, implements, import, in, indexed, instanceof, interface, internal, is, length, library, log0, log1, log2, log3, log4, memory, modifier, new, payable, pragma, private, protected, public, pure, push, require, return, returns, revert, selfdestruct, send, solidity, storage, struct, suicide, super, switch, then, this, throw, transfer, true, try, typeof, using, value, view, while, with, addmod, ecrecover, keccak256, mulmod, ripemd160, sha256, sha3}, 
	keywordstyle=[1]\color{blue}\bfseries,
	keywords=[2]{address, bool, byte, bytes, bytes1, bytes2, bytes3, bytes4, bytes5, bytes6, bytes7, bytes8, bytes9, bytes10, bytes11, bytes12, bytes13, bytes14, bytes15, bytes16, bytes17, bytes18, bytes19, bytes20, bytes21, bytes22, bytes23, bytes24, bytes25, bytes26, bytes27, bytes28, bytes29, bytes30, bytes31, bytes32, enum, int, int8, int16, int24, int32, int40, int48, int56, int64, int72, int80, int88, int96, int104, int112, int120, int128, int136, int144, int152, int160, int168, int176, int184, int192, int200, int208, int216, int224, int232, int240, int248, int256, mapping, string, uint, uint8, uint16, uint24, uint32, uint40, uint48, uint56, uint64, uint72, uint80, uint88, uint96, uint104, uint112, uint120, uint128, uint136, uint144, uint152, uint160, uint168, uint176, uint184, uint192, uint200, uint208, uint216, uint224, uint232, uint240, uint248, uint256, var, void, ether, finney, szabo, wei, days, hours, minutes, seconds, weeks, years},	
	keywordstyle=[2]\color{teal}\bfseries,
	keywords=[3]{block, blockhash, coinbase, difficulty, gaslimit, number, timestamp, msg, data, gas, sender, sig, value, now, tx, gasprice, origin},	
	keywordstyle=[3]\color{violet}\bfseries,
	identifierstyle=\color{black},
	sensitive=false,
	comment=[l]{//},
	morecomment=[s]{/*}{*/},
	commentstyle=\color{gray}\ttfamily,
	stringstyle=\color{red}\ttfamily,
	morestring=[b]',
	morestring=[b]"
}